\documentclass[10pt,onecolumn,showkeys  ,secnumarabic, nobibnotes,  nofootinbib, aps,prd]{revtex4}%
\usepackage[utf8]{inputenc}
\usepackage{amsmath}
\usepackage{amssymb}
\usepackage{amsfonts}
\usepackage{graphicx}
\usepackage{hyperref}%
\begin{document}
%\Title{Kinematics of the Milky Way from the Statistical Analysis of the Gaia Data Release 3}
\title{Kinematics of the Milky Way from the Statistical Analysis of the Gaia Data Release 3}
% Author Orchid ID: enter ID or remove command
%\newcommand{\orcidauthorA}{0000-0002-8296-2128} % Add \orcidA{} behind the author's name
%\newcommand{\orcidauthorB}{0000-0002-1341-7783} % Add \orcidB{} behind the author's name
%\Author{Petr Zavada *\href{https://orcid.org/0000-0002-8296-2128}{\orcidicon} and Karel P\'{\i}\v{s}ka \href{https://orcid.org/0000-0002-1341-7783}{\orcidicon}}
%\AuthorNames{Petr Zavada, Karel P\'{\i}\v{s}ka}

\author{Petr Zavada}
\author{Karel P\'{\i}\v{s}ka}

\begin{abstract}
The aim of the analysis of data from the Gaia Space Observatory is to obtain
kinematic parameters of the collective motion of stars in a part of our
galaxy. This research is based on a statistical analysis of the motion of
{$55,038,539$} %MDPI: please check if this should be 55,038,539 PZ: ok, corrected.
stars selected in different directions from the Sun up to a
distance of 3--6 kpc. We developed statistical methods
for the analysis working with input data represented by the full astrometric solution (five parameters). Using the proposed statistical
methods, we obtained the local velocity of the Sun $\left(  U_{\odot}%
,V_{\odot},W_{\odot}\right)  =(9.58,16.25,7.33)\pm(0.05,0.04,0.02)_{stat}%
\pm(0.7,0.9,0.1)_{syst}$ km/s and the rotation velocity of the galaxy at
different radii. For the Sun's orbit radius, we obtained the velocity
of the galaxy rotation$\ V_{c}\approx234\pm4$ km/s. Collective rotation slows
down in the region under study linearly with distance from the disk plane:
$\Delta V/\Delta Z\cong33.5~\mathrm{km\,s}^{-1}\mathrm{kpc}^{-1}$. We showed that the different kinematic characteristics and distributions, which
depend on the position in the galaxy, can be well described in the studied 3D
region by a simple Monte Carlo simulation model, representing an
axisymmetric approximation of the galaxy kinematics. The optimal values of the
six free parameters were tuned by comparison with the data.
\end{abstract}

% Keywords
%\keyword{Milky Way; Galactic kinematics; Statistical methods; Monte Carlo simulation}
%\keywords{Methods: statistical -- Methods: data analysis -- Surveys -- Astrometry -- Stars: binaries -- Gravitational lensing: micro}

% Affiliations / Addresses (Add [1] after \address if there is only one affiliation.)
%\address{Institute of Physics of the Czech Academy of Sciences, Na Slovance 2, 182 21 Prague, Czech Republic}
\affiliation{Institute of Physics of the Czech Academy of Sciences, Na Slovance 2, 182 21 Prague 8, Czech Republic}
\email{zavada@fzu.cz}

\maketitle

%%%%%%%%%%%%%%%%%%%%%%%%%%%%%%%%%%%%%%%%%%

\section{Introduction}
\label{intro}Our galaxy, the Milky Way (MW), is a unique laboratory for
gravity research and for understanding the formation and evolution of
galaxies. In recent years, the Gaia Space Observatory has acquired a huge
amount of precise astrometric, photometric, and spectroscopic data on stars in
the {MW} %MDPI:  References should be numbered in order of appearance from 1--50. please rearrange all the references to appear in numerical order PZ: references are rearranged.
\cite{ga3}. The analysis of these data has been the subject of many
thousands of publications.

The Gaia DR3 catalogue provides very rich input data for creating a kinematic map of the MW, much more detailed than previously possible. In
general, it encompasses various structures on different scales, from the orbiting
of small gravitationally bound systems, binaries, and multiple-bound systems \cite{za3}
to the streaming motions of stellar fields in spiral arms~\cite{ant16,ant18}
with various turbulences and fluctuations and to the collective rotation of the
whole galactic disk with the galactic halo.

The nature of the rotation suggests the presence of dark matter, which
generates a substantial part of the MW mass, see
\cite{bha14,eil19,mro19,rei14,bur13,chr20,hua16,jia21,weg19,abl17} and the
overview \cite{sof20}. Recent studies of MW kinematics have shown accurate
results on the MW rotation represented by the rotation curve (RC) defined as
the dependence of the orbital velocity on the radius. These curves are the
basic input for models accounting for the presence of dark matter. This is why
these two topics often appear together in papers.

Along with gravity, the formation and evolution of the stars themselves are
also controlled by the forces of the microworld (strong, electroweak---unified
electromagnetic+weak) based on a well-verified standard model. The nature and
origin of dark matter at the microscopic level have not yet been explained.

Very important topics concern the detailed mapping of many kinematic
substructures beyond axial symmetry
\cite{gaia2023,ga1,kaw18,lop19,ram18,wan22,wan20,kus17,kaw19,tia17}. An
overview of the integrated, structural, and kinematic parameters of the galaxy
is given in \cite{bla16}.

The first goal of our study is to use statistical methods to determine the local velocity of the Sun and to analyse the collective orbital motion of stars in the galactic plane and beyond. Because the methods used do not require knowledge of radial velocities, we can access larger samples of stars
\cite{ant17,mik22}. The second goal is to obtain the 3D axisymmetric approximation of the kinematic MW image generated by averaging local asymmetric substructures. This image can be useful in determining the scale of local asymmetries or as input for validating axisymmetric models.

In Section~\ref{methodology}, we describe our methodology and define the basic
concepts and quantities we will work with. First, we define transformation
relations between galactic and Galactocentric reference frames. The first
system is used to acquire and present Gaia data, while the second is suitable
for simulation and data interpretation. Then, the definition of the
axisymmetric Monte Carlo model in the Galactocentric reference frame follows.
The model is based on a triple (partially asymmetric) Gaussian distribution,
which depends on the distance from the galactic plane and is defined by six free
parameters. Next, we define the sky sectors to be analysed and the cuts for
selecting accurate data. In Section~\ref{results}, we present our methods together
with the results obtained from velocity distributions in the different
sectors of the sky. Analyses of these distributions provide precise
results concerning the local and orbital velocity of the Sun (Section~\ref{lovesu}), the different RC representations (\mbox{Section~\ref{velcur}}), and the setting of the free parameters of the simulation model (Section~\ref{colmot}). This is followed by a comparison of the simulation with all relevant distributions (\mbox{Section~\ref{cosida}}). The obtained results and their agreement with the simulation model are discussed in more detail in Section~\ref{discussion}. The discussion includes a comparison of the results obtained with other available data.

\section{Methodology and Definitions}
\label{methodology}
\subsection{Reference Frames}
\begin{figure}[h]%[H]%[t]
\includegraphics[width=95mm]{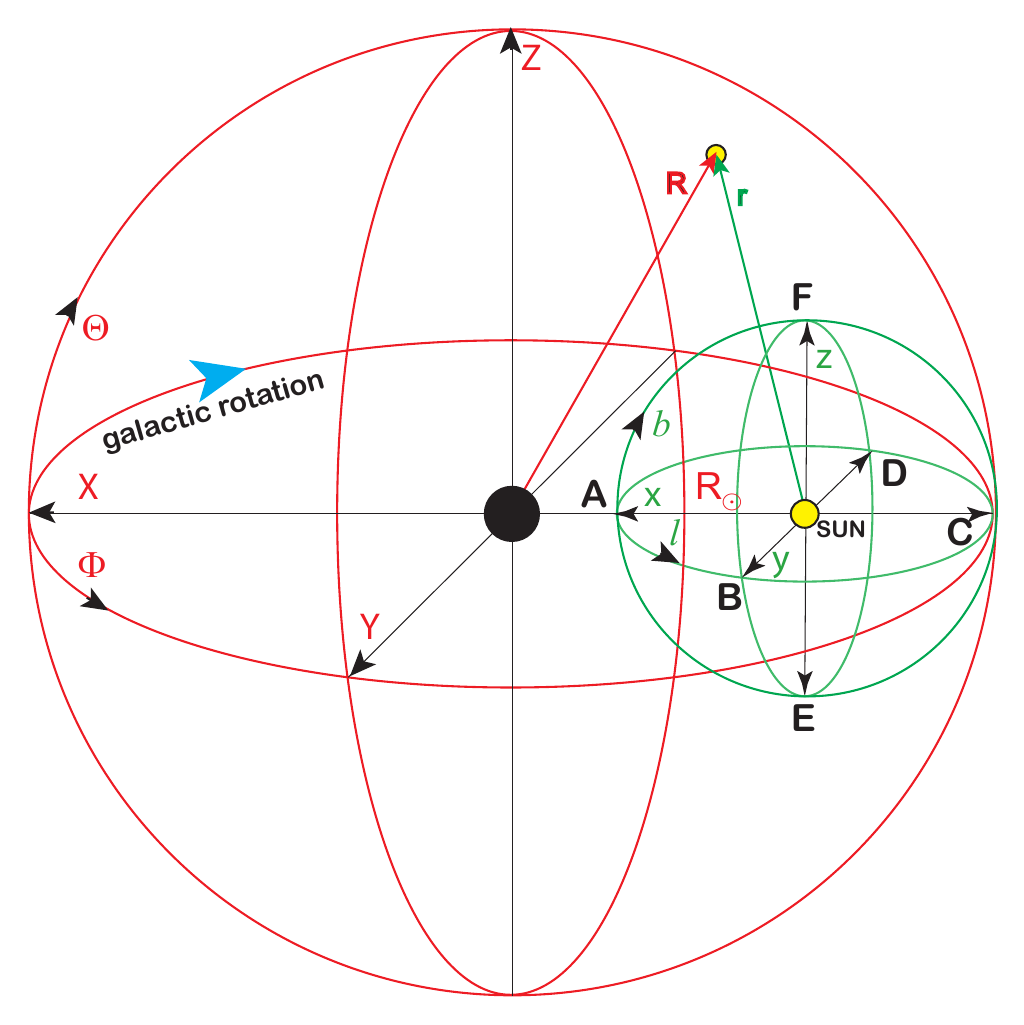}\caption{Galactic (green) and
Galactocentric (red) reference frames. The letters A--F indicate the directions of the primary sectors of our analysis, see Table~\ref{Tab1}.}%
\label{fg1}%
\end{figure}

Positions of sources in Gaia data are represented in angular galactic
coordinates: longitude ($l$) and latitude ($b$). Using parallax, we
can also define the distance $r$ of the source from the Sun. For our analysis,
representation in Galactocentric coordinates will also be important. The relation
between both reference frames is illustrated in Figure~\ref{fg1}. The axes of the galactic and Galactocentric reference frames are
defined as%
\vspace{-6pt}
\begin{align}
x &  =r\cos l\cos b,\ \ \ \ \ y=r\sin l\cos b,\ \ \ \ \ z=r\sin b,\label{me2t}%
\\
X &  =R\cos\Phi\cos\Theta,\ Y=R\sin\Phi\cos\Theta,\ Z=R\sin\Theta.\nonumber
\end{align}

\noindent For clarity, we will use the following convention throughout the paper: positions, velocities, and their coordinates are denoted in lowercase (uppercase) in the galactic (Galactocentric) reference frame. So, the direction $x$ points to the centre of the galaxy and the direction $y$
is the direction of the MW rotation. The corresponding coordinates are
related as%
\vspace{-6pt}
\begin{equation}
X=x-R_{\odot};\qquad Y=y;\qquad Z=z.\label{me2u}%
\end{equation}
Radius $R_{\odot}$ represents the distance of the Sun from the galactic
centre. The values obtained in the measurements are in the interval 7.1--8.92 kpc \cite{bla16}. For our analysis, we used the recent value $R_{\odot}%
\approx8.178$ kpc \cite{gra19}.  The small parameter $Z_{\odot}\approx0.0208$
kpc \cite{ben19} defining the position of the Sun above the galactic plane is
neglected in our analysis. The validity of this approximation will be
commented on in Section~\ref{colmot}.
\vspace{-9pt}

Then, the transformation between spherical coordinates $\left(  r,l,b\right)
\rightarrow\left(  R,\Phi,\Theta\right)  $ of both the frames is defined
by equations%
\begin{align}
R  &  =\sqrt{r^{2}+R_{\odot}^{2}-2R_{\odot}x},\qquad\Theta=\arcsin\frac{z}%
{R},\label{me1}\\
\Phi\left(  x>R_{\odot}\right)   &  =\arcsin\frac{y}{R\cos\Theta},\qquad
\Phi\left(  x\leq R_{\odot}\right)  =\pi-\arcsin\frac{y}{R\cos\Theta}\nonumber
\end{align}
and inversely $\left(  R,\Phi,\Theta\right)  \rightarrow$ $\left(
r,l,b\right):  $%

\begin{align}
r  &  =\sqrt{R^{2}+R_{\odot}^{2}+2R_{\odot}X},\qquad b=\arcsin\frac{Z}%
{r}\label{me2}\\
l\left(  X>R_{\odot}\right)   &  =\pi-\arcsin\frac{Y}{r\cos b},\qquad l\left(
X\leq R_{\odot}\right)  =\arcsin\frac{Y}{r\cos b},\nonumber
\end{align}
where $x,y,z,\ X,Y,Z$ are defined by (\ref{me2t}). We will need these
transformations to model and simulate the motion of the stars in
Sections~\ref{simul} and \ref{cosida}. The spherical coordinates $\left(  R,\Phi,\Theta\right)  $
are simply related to the cylindrical coordinates $\left(  R_{c}%
,\Phi,Z\right)  $:%
\vspace{-6pt}
\begin{equation}
R_{c}=R\cos\Theta;\qquad\ Z=R\sin\Theta. \label{me2a}%
\end{equation}

Next, we will need
vectors forming the local orthonormal bases in both reference frames:%
\vspace{-6pt}
\begin{align}
\mathbf{N}_{\Phi}=\left(  -\sin\Phi,\cos\Phi,0\right)  ,\qquad\mathbf{N}%
_{\Theta}=\left(  -\cos\Phi\sin\Theta,-\sin\Phi\sin\Theta,\cos\Theta\right)
,\label{me3a}\\
\mathbf{N}_{R}=\left(  \cos\Phi\cos\Theta,\sin\Phi\cos\Theta,\sin
\Theta\right)  ,\nonumber\\
\mathbf{n}_{l}=\left(  -\sin l,\cos l,0\right)  ,\qquad\mathbf{n}_{b}=\left(
-\cos l\sin b,-\sin l\sin b,\cos b\right)  ,\label{me3b}\\
\mathbf{n}_{r}=\left(  -\cos l\cos b,-\sin l\cos b,\sin b\right),  \nonumber
\end{align}
where the vectors $\mathbf{N}_{\alpha}$ and $\mathbf{n}_{\alpha}$ define
directions of increasing coordinates $\alpha=\Phi,\Theta,R,\ l,b,r$. For
example, $-\mathbf{N}_{\Phi}$ is the direction of the MW rotation at any point.

The velocity of a star at the point $\mathbf{R}$\ of a Galactocentric frame can
be split into two components:%
\vspace{-6pt}
\begin{equation}
\mathbf{V}\left(  \mathbf{R}\right)  =\mathbf{V}_{G}\left(  \mathbf{R}\right)
+\Delta\mathbf{V}\left(  \mathbf{R}\right)  ;\qquad \mathbf{V}_{G}\left(
\mathbf{R}\right)  =-V_{G}(R,Z)\mathbf{N}_{\Phi},\label{me9c}%
\end{equation}
where $V_{G}(R,Z)>0$ is the average (over $\Phi$) rotation velocity at the radius $R$ and distance $Z$ from the galactic plane. $\Delta
\mathbf{V}$\ is the deviation from the average
$\mathbf{V}_{G}\left(  \mathbf{R}\right)  $, which represents the velocity of the local rest frame and can be approximated by the average velocity of the stars in some neighbourhood of $\mathbf{R}$:
\vspace{-6pt}
\begin{equation}
\mathbf{V}_{G}\left(  \mathbf{R}\right)  =\left\langle \mathbf{V}\left(
\mathbf{R}\right)  \right\rangle ;\qquad\left\langle \Delta\mathbf{V}\left(
\mathbf{R}\right)  \right\rangle =0.\label{me9d}%
\end{equation}
The quality of the approximation may depend on the choice of sources and the
size and shape of the defined neighbourhood. In Section~\ref{lovesu}, this issue will be
discussed in more detail for $\mathbf{V}_{G}\left(  \mathbf{R}_{\odot}\right)
$, which is the velocity of the local standard rest frame (LSR) \cite{sch10}.

The proper motion of the stars in Gaia data is represented by the vector%
\vspace{-6pt}
\begin{equation}
{\mu}_{ICRS}=\left(  \mu_{\alpha}^{\ast},\mu_{\delta}\right)
;\qquad\mu_{\alpha}^{\ast}\equiv\mu_{\alpha}\cos\delta,\label{me3}%
\end{equation}
whose components are angular velocities in directions of the right ascension
and declination in the ICRS. For our analysis, we will prefer the
representation of proper motion in the galactic reference frame:%
\vspace{-6pt}
\begin{equation}
{\mu}_{gal}=\left(  \mu_{l}^{\ast},\mu_{b}\right)  ,\qquad\mu_{l}%
^{\ast}\equiv\mu_{l}\cos b,\label{me4}%
\end{equation}
which is obtained by the transformation given in \cite{ga4}.
Then, we will work with transverse 2D velocity defined as%
\vspace{-6pt}
\begin{equation}
\mathbf{v}_{gal}\mathbf{=}\left(  v_{l},v_{b}\right)  [\mathrm{km/s}%
]\mathbf{=}4.7406\,\,r[\mathrm{kpc}]\,\left(  \mu_{l}^{\ast},\mu_{b}\right)
[\mathrm{mas/year}], \label{me8}%
\end{equation}
where $v_{l},v_{b}$ are velocity components\ in directions of increasing
latitude $\left(  l\right)  $ and longitude $\left(  b\right)  ,$ and the
distance $r$ is obtained from the parallax:
\vspace{-6pt}
\begin{equation}
r[\mathrm{kpc}]=\frac{1}{p[\mathrm{mas}]-\mathrm{ZP}};\qquad\mathrm{ZP}%
=-0.017\mathrm{mas}, \label{me9}%
\end{equation}
where ZP is a global zero point taken from \cite{gaia2021}. The ZP correction
reduces velocity only by $\lesssim2\%$ in our selected data. The analysis and
discussion of distance extraction from parallax is thoroughly conducted in
\cite{bai15,gaia2018,ga2}.

In Section~\ref{lovesu}, we will study the dependence of the mean values
$\left\langle v_{l}\right\rangle ,\left\langle v_{b}\right\rangle $ on the
distance and direction from the Sun. From these curves, we will determine the
local velocity of the Sun. In the rest of Section~\ref{results}, we will work with
slightly modified velocities $\left\langle v_{l}\right\rangle ,\left\langle
v_{b}\right\rangle $ and corresponding dispersions that are related to the LSR. In this reference frame, the dependence on the Sun's local motion is eliminated.

\subsection{Simulation Model of Stellar Velocities}
\label{simul}
The velocity distributions will be compared with a simple probabilistic
Monte Carlo model. The model generates the velocity of a star in the
Galactocentric reference frame (Figure~\ref{fg1}):
\begin{equation}
\mathbf{V}^{gen}\mathbf{=}\left(  V_{\Phi}-V_{0}\left(  R\right)  \right)
\mathbf{N}_{\Phi}+V_{\Theta}\mathbf{N}_{\Theta}+V_{R}\mathbf{N}_{R},
\label{ge1}%
\end{equation}
where $V_{\Phi},V_{\Theta}$, and $V_{R}$ are its components generated in the
local reference frame defined by the basis (\ref{me3a}). Velocity
$V_{0}\left(  R\right)  >0$ is defined as an average of orbital velocity at
the galactic plane and radius $R$:%
\begin{equation}
V_{0}\left(  R\right)  =V_{G}(R,0), \label{ge2}%
\end{equation}
which is also our definition of the RC.\ This definition is based on direct
measurements of orbital velocities in selected sectors of the MW disk, so the
results obtained may differ from a global RC calculated from Jeans modelling
\cite{jea1915} assuming an axisymmetric gravitational potential of the MW. Our
definition reflects the collective orbital velocity rather than the velocity
of a single star or a test particle \cite{eil19}.

As we shall see, a very good agreement with the observed curves and
distributions provides the simulation model based on the multinormal distribution:%
\vspace{-6pt}
\begin{equation}
P(V_{\Phi}\mathbf{,}V_{\Theta},V_{R};R,Z\mathbf{)\sim}\exp\left(
-\frac{V_{\Phi}^{2}}{2\sigma_{\Phi}^{2}}-\frac{V_{\Theta}^{2}}{2\sigma
_{\Theta}^{2}}-\frac{V_{R}^{2}}{2\sigma_{R}^{2}}\right)  , \label{ge3}%
\end{equation}
with partial asymmetry---a different $\sigma_{\Phi}$ for positive and negative
$V_{\Phi}$. The dependence on $R$ and $Z$ is absorbed in the standard
deviations $\sigma_{\alpha}(R,Z)$. This important dependence will be discussed in more detail in Section~\ref{colmot}. For comparison with the data, the velocities simulated in the Galactocentric system will be transformed into the galactic system as described in Section~\ref{cosida}.

\subsection{Data Set}

\begin{figure}[h]%[H]%[t]
\includegraphics[width=135mm]{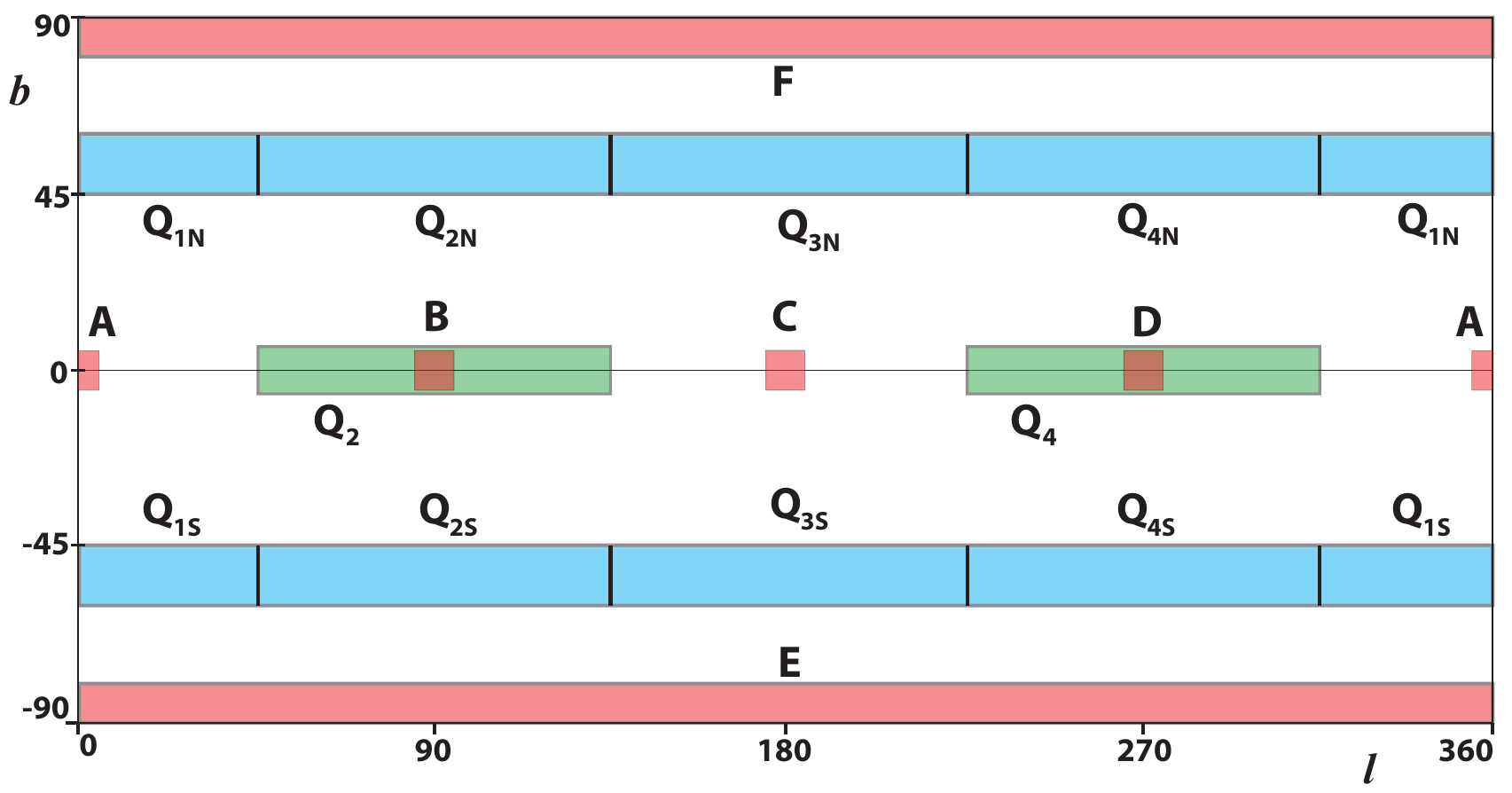}\caption{{MW} %MDPI: Please change the hyphen (-) into a minus sign ($-$, "U+2212"). e.g., "-1" should be "$-$1". PZ: ok, but fg2.pdf is produced differently
sectors used for
analysis. Sectors B and D are subsets of sectors Q$_{2}$ and Q$_{4}$. Units:
$b,l$ [deg].}%
\label{fg2}%
\end{figure}

The data sectors of the sky used for analysis are defined in
Figure~\ref{fg2} and Table~\ref{Tab1}.
The narrow sectors A--F are defined by basic directions (from the
Sun, see Figure~\ref{fg1}): towards the centre of the galaxy (A), along the
direction of galaxy rotation (B), and their opposites (C,D). The
perpendiculars to the galactic disk define sectors E and F. The analysis is
extended by other sectors: Q$_{\alpha}$---in the galactic disk and
Q$_{\alpha\beta}$---outside the disk.
We only accept sources with a positive
parallax and limited distance $r<6$ kpc that satisfy the following condition:%
\vspace{-6pt}
\begin{equation}
\frac{\Delta v}{v}\leq\eta;\qquad v=\left\vert \mathbf{v}_{gal}\right\vert
=\left\vert \mathbf{v}_{ICRS}\right\vert ,\label{me9b}%
\end{equation}
whereby we set cuts for different sectors:%
\vspace{-6pt}
\[%
\begin{tabular}
[c]{cccc}
& A & B-D & E,F,Q\\\hline
$\eta$ & 0.05 & 0.2 & 0.3
\end{tabular}
\
\]
In sectors from which we will determine our resulting velocities in
Table~\ref{Tab2}, we use a more strict cut. Sector A contains regions of very
high and inhomogeneous density and only a very safe cut $\eta$ gives
consistent results compatible with the other sectors. We verified that the narrowing of these cuts does not systematically affect the curves, from which the resulting parameters are obtained. The error $\Delta
v$ is estimated as%
\begin{equation}
\left(  \Delta v\right)  ^{2}=\left(  \frac{\partial v}{\partial p}\Delta
p\right)  ^{2}+\left(  \frac{\partial v}{\partial\mu_{\alpha}^{\ast}}\Delta
\mu_{\alpha}^{\ast}\right)  ^{2}+\left(  \frac{\partial v}{\partial\mu
_{\delta}}\Delta\mu_{\delta}\right)  ^{2}.\label{me9e}%
\end{equation}

\noindent Using Equation~(\ref{me8}) expressed in ICRS and relation (\ref{me9}), we obtain the following ratio:
\begin{equation}
\frac{\Delta v}{v}\approx\sqrt{\left(  \frac{\Delta p}{p}\right)  ^{2}%
+\frac{\left(  \mu_{\alpha}^{\ast}\Delta\mu_{\alpha}^{\ast}\right)
^{2}+\left(  \mu_{\delta}\Delta\mu_{\delta}\right)  ^{2}}{\mu^{4}}};\qquad
\mu=\sqrt{\mu_{\alpha}^{\ast2}+\mu_{\delta}^{2}},\label{me9f}%
\end{equation}
which is calculated from the Gaia data. The resulting numbers of accepted
sources in the respective sectors are shown in the same table. Most of our calculations focus on mean values, which means that the resulting errors are much smaller than the errors of individual entries.

\vspace{-6pt}

\vspace{-9pt}
\begin{table}[h]%[H]%[ptb]
\caption{Number %MDPI: We moved Table 1 after where it was first mentioned. Please confirm. PZ: Please put it back into the section Data set--%MDPI: done, we put it back to its original seciton. We moved Figure 2 and Table 1 after their discussion, please check and confirm the place.
of stars in the analysed MW sectors: N$_{tot}$---all stars,
n$_{cut}$---selected stars satisfying cut (\ref{me9b}), and $\left\langle
\mathrm{r}\right\rangle $---mean distance of selected stars. }%
\label{Tab1}%
%\begin{adjustwidth}{-\extralength}{0cm}
\setlength{\tabcolsep}{2.76mm}
\begin{tabular}{ccccccccc}
\hline
& A & B & C & D & E & F & Q$_{2}$ & Q$_{4}$\\\hline
$b[\deg]$ & \multicolumn{4}{c}{$\left\langle -5,5\right\rangle $} &
$\left\langle -90,-80\right\rangle $ & $\left\langle 80,90\right\rangle $ &
\multicolumn{2}{c}{$\left\langle -5,5\right\rangle $}\\
$l[\deg]$ & $\left\langle -5,5\right\rangle $ & $\left\langle
85,95\right\rangle $ & $\left\langle 175,185\right\rangle $ & $\left\langle
265,275\right\rangle $ & \multicolumn{2}{c}{$\left\langle 0,360\right\rangle
$} & $\left\langle 45.135\right\rangle $ & $\left\langle 225,315\right\rangle
$\\
$\left\langle \mathrm{r}\right\rangle [\mathrm{kpc}]$ & 1.02 & 1.89 & 1.63 &
1.84 & 0.90 & 0.93 & 1.93 & 2.08\\
n$_{cut}$ & 253,546 & 1,612,100 & 609,593 & 1,233,495 & 314,515 & 283,906 & 18,968,282 &
24,584,392\\
N$_{tot}$ & 22,704,200 & 6,646,423 & 2,498,909 & 5,422,928 &  573,436 & 540,330 & 70,260,517 & 85,849,109\\
\hline
H
& Q$_{1S}$ & Q$_{2S}$ & Q$_{3S}$ & Q$_{4S}$ & Q$_{1N}$ & Q$_{2N}$ & Q$_{3N}$ &
Q$_{4N}$\\\hline
\multicolumn{1}{c}{$b[\deg]$} & \multicolumn{4}{c}{$\left\langle
-60,-45\right\rangle $} & \multicolumn{4}{c}{$\left\langle 45,60\right\rangle
$}\\
\multicolumn{1}{c}{$l[\deg]$} & $\left\langle -45,45\right\rangle $ &
$\left\langle 45,135\right\rangle $ & $\left\langle 135.225\right\rangle $ &
$\left\langle 225,315\right\rangle $ & $\left\langle -45,45\right\rangle $ &
$\left\langle 45,135\right\rangle $ & $\left\langle 135.225\right\rangle $ &
$\left\langle 225,315\right\rangle $\\
$\left\langle \mathrm{r}\right\rangle [\mathrm{kpc}]$ & 0.89 & 0.75 & 0.72 &
0.87 & 0.89 & 0.87 & 0.74 & 0.79\\
\multicolumn{1}{c}{n$_{cut}$} & 1,646,264 & 1,161,311 & 995,955 & 1,475,289 &
1,378,017 & 1,337,495 & 977,358 & 1,052,616\\
\multicolumn{1}{c}{N$_{tot}$} & 3,331,201  &  2,164,099 &1,716,957 & 2,640,374 & 2,973,013 & 2,259,886 & 1,688,028 &2,153,065\\
\hline
\end{tabular}

%\end{adjustwidth}
\end{table}

\section{Methods and Results}
\label{results}
\begin{figure}[h]%[H]
%\begin{adjustwidth}{-\extralength}{0cm}
\centering\includegraphics[width=170mm]{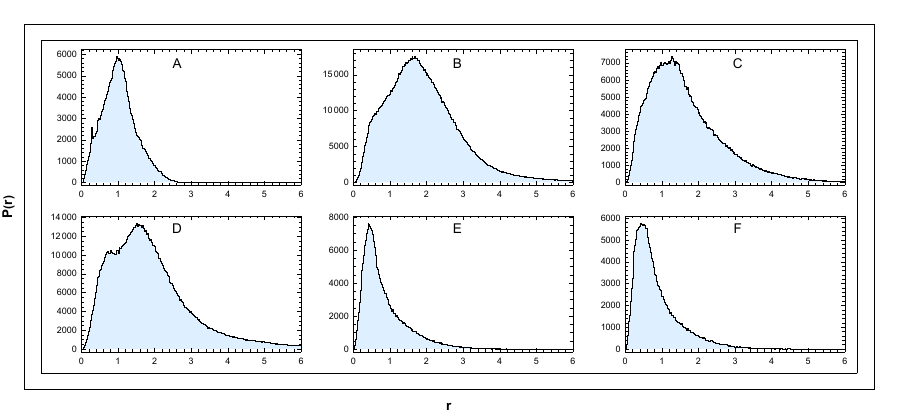}
%\end{adjustwidth}
\caption{{Distribution} %MDPI: Please use commas to separate thousands for numbers with five or more digits (not four digits) in the picture, e.g., "10000" should be "10,000". PZ: this is probably related to Table 2, where commas are inserted...
of
distances in the sectors (\textbf{A}--\textbf{F}). Unit: r[kpc]. Binning: 0.024~kpc.}%
\label{fg3}%
\end{figure}

In Figure~\ref{fg3}, we show the distribution of star distances in the data sectors A--F
defined above. The distance of most of them is $r\lesssim$ 3--6~kpc.
Dependencies of mean velocities $\left\langle v_{l}\right\rangle ,\left\langle
v_{b}\right\rangle ,\left\langle v_{gal}\right\rangle$ and related standard
deviations on distance $r$ are shown in the figures that follow. What
information can be extracted from them?

\subsection{Velocity of the Sun}
\label{lovesu} 
\begin{figure}[h]%[H]
%\begin{adjustwidth}{-\extralength}{0cm}
\centering\includegraphics[width=170mm]{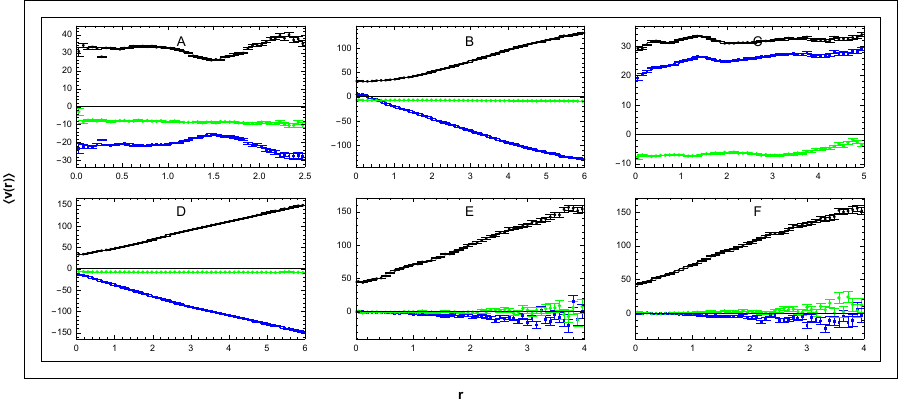}
%\end{adjustwidth}
\caption{Dependence of mean
velocity ($v_{l}$---blue, $v_{b}$---green, and $v_{gal}$---black) on distance $r$ in
the galactic reference frame in sectors (\textbf{A}--\textbf{F}). Units: r[kpc],
v[km/s]. }%
\label{fg4}%
\end{figure}

According to (\ref{me9c}) and (\ref{ge2}), the Sun's velocity
can be decomposed as%

\begin{equation}
\mathbf{V}_{\odot}=\mathbf{V}_{G}\left(  \mathbf{R}_{\odot}\right)
+\Delta\mathbf{V}_{\odot};\quad\mathbf{V}_{G}\left(  \mathbf{R}_{\odot
}\right)  =-V_{0}(R_{\odot})\mathbf{N}_{\Phi},\label{r1}%
\end{equation}
where $\mathbf{V}_{G}\left(  \mathbf{R}_{\odot}\right)  $ is the LSR velocity
and $\Delta\mathbf{V}_{\odot}$ is the Sun's velocity in the LSR. How
can these velocities be determined? In the galactic reference frame, the star
velocity $\mathbf{v(r)}$ is given~as%
\begin{equation}
\mathbf{v(r)}=\mathbf{V}\left(  \mathbf{R}_{\odot}\mathbf{+r}\right)
-\mathbf{V}_{\odot}=\mathbf{V}_{G}\left(  \mathbf{R}_{\odot}\mathbf{+r}%
\right)  +\Delta\mathbf{V}\left(  \mathbf{R}_{\odot}\mathbf{+r}\right)
-\mathbf{V}_{G}\left(  \mathbf{R}_{\odot}\right)  -\Delta\mathbf{V}_{\odot
},\label{r2a}%
\end{equation}
which implies%
\begin{equation}
\left\langle \left\langle \mathbf{v(r)}\right\rangle \right\rangle
=\left\langle \left\langle \mathbf{V}\left(  \mathbf{R}_{\odot}\mathbf{+r}%
\right)  \right\rangle \right\rangle -\mathbf{V}_{G}\left(  \mathbf{R}_{\odot
}\right)  -\Delta\mathbf{V}_{\odot}.\label{r2b}%
\end{equation}
The region of averaging $\left\langle \left\langle \mathbf{...}\right\rangle
\right\rangle $ $\left(  \text{over }\mathbf{r}\right)  $ should be reasonably
wide to suppress the influence of local fluctuations on the obtained average,
and then the first two terms cancel, and for the local Sun's velocity, we obtain%
\vspace{-6pt}
\begin{equation}
\Delta\mathbf{V}_{\odot}=-\left\langle \left\langle \mathbf{v(r)}\right\rangle
\right\rangle .\label{r2c}%
\end{equation}
For its calculation, we use the projections of galactic velocities (\ref{me8})
in sectors A--D shown in Figure~\ref{fg4} together with the relations
\vspace{-6pt}
\begin{equation}
v_{l}\mathbf{(r)}=\mathbf{v(r).n}_{l},\mathbf{\qquad}v_{b}\mathbf{(r)}%
=\mathbf{v(r).n}_{b},\label{r2d}%
\end{equation}
where the vectors of local orthonormal basis $\mathbf{n}_{\alpha}%
$\ are defined in (\ref{me3b}). In the sectors considered, we obtain

(1) $\Delta V_{z\odot}$ in the sectors A--D

\begin{figure}[h]%[H]
%\begin{adjustwidth}{-\extralength}{0cm}
\includegraphics[width=135mm]{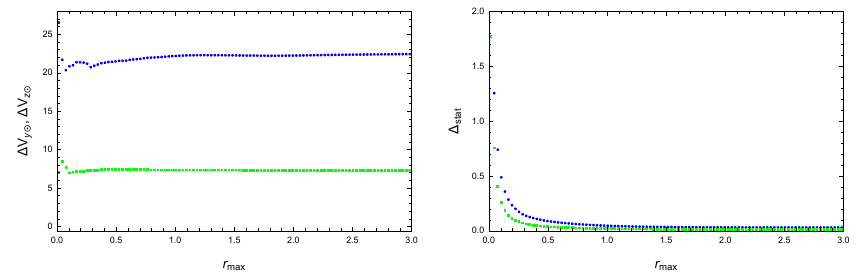}\caption{Velocities
$\Delta V_{y\odot}$ (blue%MDPI: Please confirm if the italics is unnecessary and can be removed. The following highlights are the same. PZ: italics removed.
) and  $\Delta V_{z\odot}$ (green) as the functions
of r$_{\max}$~(\textbf{left}) with corresponding statistical errors (\textbf{right}). Units: r$_{\max}$[kpc],  $\Delta V_{\alpha\odot}$[km/s].}%
\label{fg4VW}%
%\end{adjustwidth}
\end{figure}

Since in these sectors we have $\mathbf{n}_{b}=\left(  0,0,1\right)  ,$ so we
can identify $v_{z}=v_{b}$. The mean values $\left\langle v_{b}\right\rangle $
depending on distance $r$ are for individual sectors shown in Figure~\ref{fg4}.
The velocity $\Delta V_{z\odot}$ is the average:%
\begin{equation}
\Delta V_{z\odot}\left(  r_{\max}\right)  =-\left\langle \left\langle
v_{z}\right\rangle \right\rangle _{\Omega }, \label{r3}%
\end{equation}
where $\Omega $ means the region of averaging, which are the stars in the sectors A--D
and radius $r<$ $r_{\max}$. The resulting curve is shown in Figure~\ref{fg4VW}.

%\vspace{-6pt}

(2) $\Delta V_{y\odot}$ in the sectors A and C

In these sectors we have $\mathbf{n}_{l}=\left(  0,\pm1,0\right)  $, so we
identify $v_{y}=+v_{l}$ in the A and $v_{y}=-v_{l}$ in the C sector. The
corresponding curves $\left\langle v_{l}\right\rangle $ are in panels A,C
(Figure~\ref{fg4}) and curve%
\begin{equation}
\Delta V_{y\odot}\left(  r_{\max}\right)  =-\frac{1}{2}\left(  \left\langle
\left\langle v_{y}\right\rangle \right\rangle _{A}+\left\langle \left\langle
v_{y}\right\rangle \right\rangle _{C}\right)  \label{r6}%
\end{equation}
is in Figure~\ref{fg4VW}. The region of averaging are sectors A and C up to
the distance $r_{\max}.$

(3) $\Delta V_{x\odot}$ and $V_{0}(R_{0})$ in the sectors B and D

\begin{figure}[h]%[H]
%\begin{adjustwidth}{-\extralength}{0cm}
\hspace{-17pt}\includegraphics[width=135mm]{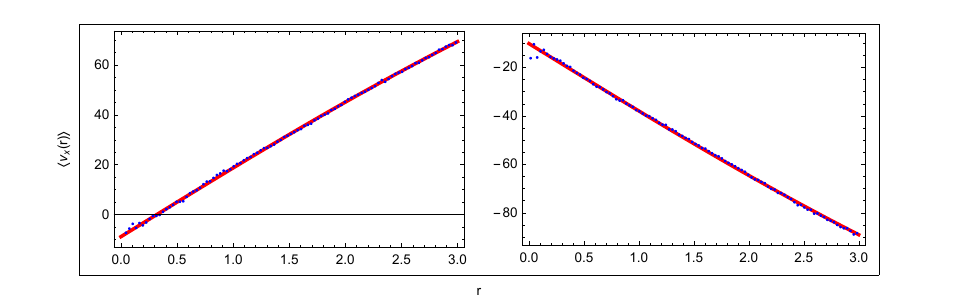}\caption{Curves (\ref{r5a}):
data (blue%MDPI: Please confirm if the italics is unnecessary and can be removed. The following highlights are the same. PZ: italics removed
) and fit (red). Units: r[kpc],
$v_{x}$[km/s].}%
\label{fg4U}%
%\end{adjustwidth}
\end{figure}

Here we have $\mathbf{n}_{l}=\left(  \mp1,0,0\right)  $, so we can identify
$v_{x}=-v_{l}$ in sector B and $v_{x}=+v_{l}$ in the D sector. In panels B,D in
Figure~\ref{fg4}, we show corresponding $\left\langle v_{l}\right\rangle
$\ curves. For small $r$, we have $\Delta V_{x\odot}=-\left\langle
v_{x}\right\rangle $, which implies for B,D sectors,%
\vspace{-6pt}
\begin{equation}
\Delta V_{x\odot}=\pm\left\langle v_{l}\right\rangle .\label{r4}%
\end{equation}
but what is the reason for the steep slope of $\left\langle v_{l}\right\rangle
$ as $r$ increases? In the considered sectors, the mean value $\left\langle
v_{l}\right\rangle $ includes a significant contribution from the collective
rotation velocity $\mathbf{V}_{G}\left(  \mathbf{R}_{\odot}\mathbf{+r}\right)
$ proportional to $r$, as explained below ({Figure 9} %MDPI: Wrong citation number:  Figure 9 was mentioned before figure 6. Please revise figure citation order. PZ: order is correct, please do not change. %\ref{fg5}  %%MDPI: according to our guideline, all figures and tables should be refered in order, in this case, we need to remove the link of figure 9 and replace it with the text, please confirm the format change.
and Equation~(\ref{r9})).
Thus, in the B,D sectors, we have%
\begin{equation}
\left\langle v_{l}^{B}\right\rangle =\Delta V_{x\odot}-\frac{r}{R}V_{0}\left(
R\right)  ,\mathbf{\qquad\ }\left\langle v_{l}^{D}\right\rangle =-\Delta
V_{x\odot}-\frac{r}{R}V_{0}\left(  R\right)  ;\mathbf{\ \qquad}R=\sqrt
{r^{2}+R_{\odot}^{2}}\label{r5}%
\end{equation}
and correspondingly,%
\begin{equation}
\left\langle v_{x}^{B}\right\rangle =-\Delta V_{x\odot}+\frac{r}{R}%
V_{0}\left(  R\right)  ,\mathbf{\qquad\ }\left\langle v_{x}^{D}\right\rangle
=-\Delta V_{x\odot}-\frac{r}{R}V_{0}\left(  R\right)  .\label{r5a}%
\end{equation}
The two curves $\left\langle v_{l}\right\rangle $ are shown in Figure~\ref{fg4U}
together with the curve produced by fitting the free parameters
$\Delta V_{x\odot}$ and $V_{0}$ in the range $0<r<3$ kpc. Assuming that
rotation velocity is constant in these sectors, $V_{0}\left(  R\right)
\approx V_{0}(R_{\odot})$, we obtain very good agreement of the fit to the
data. In the next, we abbreviate $V_{0}(R_{\odot})$ by $V_{0}$. The obtained
velocities are%
\begin{equation}
\Delta V_{x\odot}^{B}=8.92,\mathbf{\qquad}V_{0}^{B}=227.31;\mathbf{\qquad
}\Delta V_{x\odot}^{D}=10.23,\mathbf{\qquad}V_{0}^{D}=228.86.\label{r7}%
\end{equation}
Admittedly, there is a weak dependency \cite{eil19,rei14}:
\vspace{-6pt}
\begin{equation}
V_{0}\left(  R\right)  =V_{0}-\kappa\left(  R-R_{\odot}\right)  ,\label{r7b}%
\end{equation}
however, the range $0<r<3$ kpc means that $8.178<R<8.711$, which is too
small a range for a reliable fit involving $\kappa$.

\vspace{-6pt}

In the left panel of Figure~\ref{fg4VW}, we observe nearly constant curves while
the right panel shows the dependence of corresponding statistical errors. The
range of distances (0--3~{kpc}) involves the dominant
part of analysed stars in sectors A--D. To estimate the Sun's local velocity
$\Delta V_{y\odot},\Delta V_{z\odot}$, we take the values in the middle of the
range $\left(  1.5\ \text{kpc}\right)  $. The velocities $\Delta V_{x\odot}$ and $V_{0}$ are obtained as averages of the
values in (\ref{r7}).

However, for consistent comparison with other analyses, the velocities $\Delta
V_{y\odot}$ and $V_{0}$ can require further correction based on the
calculation of asymmetric drift (see \cite{Bovy}, {Section~11}%MDPI: please check if this belongs to ref 14. PZ: Yes, it belongs, do not change..
):%
\begin{equation}
V_{a}=V_{c}-V_{0},\label{cor1}%
\end{equation}
where $V_{c}$ is the MW circular velocity at $R_{\odot}$ and $V_{0}$ is the mean
orbital velocity of the stars in the neighbourhood of the Sun. The Sun orbital
velocity $V_{\odot\Phi}$ in the Galactocentric frame can be decomposed
alternatively as%
\begin{equation}
V_{\odot\Phi}=V_{0}+\Delta V_{y\odot}=V_{c}+\Delta V_{y\odot}^{corr}%
,\label{cor2}%
\end{equation}
so%
\begin{equation}
\Delta V_{y\odot}^{corr}=\Delta V_{y\odot}-V_{a},\qquad V_{c}=V_{0}%
+V_{a}.\label{cor3}%
\end{equation}
The study \cite{li} estimates the asymmetric drift $V_{a}$\ around the Sun's
position to be about $6$~km/s. A very similar value can be extracted from
Figure 3 in the paper \cite{sch10}, where the parameter $\left\langle
v_{R}^{2}\right\rangle $ is replaced by our parameter $\sigma_{R0}^{2}=1089$
km$^{2}$s$^{-2}$, which is calculated in Table~\ref{Tab3} in Section~\ref{colmot}.
In the standard notation, we can identify%
\begin{equation}
U_{\odot}=\Delta V_{x\odot},\qquad V_{\odot}=\Delta V_{y\odot}-V_{a},\qquad
W_{\odot}=\Delta V_{z\odot},\qquad V_{c}=V_{0}+V_{a},\label{cor4}%
\end{equation}
where $\Delta V_{x\odot}$ and $V_{0}$ are the averages of the corresponding
values in (\ref{r7}). We have%
\begin{equation}
\Delta\mathbf{V}_{\odot}=\left(  \Delta V_{x\odot},\Delta V_{y\odot},\Delta
V_{z\odot}\right)  =\left(  9.58,22.25,7.33\right)  .\label{cor5}%
\end{equation}
The associated velocities $U_{\odot},V_{\odot},W_{\odot},V_{c}$, and $V_{0}$
are shown in Table~\ref{Tab2}. Their systematic uncertainties are estimated as
follows.  For
velocities $\Delta V_{y\odot},\Delta V_{z\odot}$, they are determined by the range of
variability (defined as $ \approx \left(  \max-\min\right)  $) of the curves in the
left panel in interval $0.5<r_{\max}<3$ kpc. For velocities $\Delta V_{x\odot}$ and $V_{0}$,
systematic uncertainty is estimated similarly from a set of fits in different intervals
of distances $\left(  0<r<2,3,4,5\text{ kpc}\right)  $. The relatively small
systematic uncertainty in $V_{0}$ reflects the relatively small local variations.

\begin{table}[h]%[H]%[ptb]
%\begin{center}%

\caption{Local %MDPI: Please move Table 2 after where it was first mentioned and keep ref citation order correct.
%PZ: order is ok, please do not change...%MDPI: the citation and the table should be in the same subsection, please doublecheck and confirm the place.
velocity of the Sun (with statistical errors), mean rotation velocity $V_{0}$,
circular velocity $V_{c}$, and their systematic errors $\Delta_{syst}$ along with the results of other analyses. Units: [km/s].}%
\label{Tab2}%
%\begin{adjustwidth}{-\extralength}{0cm}

\begin{tabular}{cccccccc}%
\hline
\boldmath\textbf{$U_{\odot}$} & \boldmath\textbf{$V_{\odot}$} & \boldmath\textbf{$W_{\odot}$} & \boldmath\textbf{$V_{0}$} & \boldmath\textbf{$V_{c}$} & \boldmath\textbf{$\Delta_{syst}$}
& \boldmath\textbf{$V_{c}+V_{\odot}$} & \textbf{Ref.}\\\hline
$9.58\pm0.05$ & $16.25\pm0.04$ & $7.33\pm0.02$ & $228$ &234  & $\left(
0.7,0.9,0.1,4.0,4.0\right)  $ & $251.0\pm5$ & \multicolumn{1}{l}{this work}\\
$11.1_{-0.75}^{+0.69}$ & $12.24\pm0.47$ & $7.25_{-0.36}^{+0.37}$ &  &  &
$\left(  1,2,0.5,-\right)  $ &  & \multicolumn{1}{l}{\cite{sch10}}\\
$9.12\pm0.10$ & $20.80\pm0.10$ & $7.66\pm0.08$ & $230\pm12$ &  &  & $250.8$ &
\multicolumn{1}{l}{\cite{bob16}}\\
$9.58\pm2.39$ & $10.52\pm1.96$ & $7.01\pm1.67$ &  &  &  &  &
\multicolumn{1}{l}{\cite{tia15}}\\
$10.00\pm0.36$ & $5.25\pm0.62$ & $7.17\pm0.38$ &  &  &  &  &
\multicolumn{1}{l}{\cite{deh1998}}\\
& $14.6\pm5$ &  &  & $240\pm8$ &  & $255.2\pm5.1$ &
\multicolumn{1}{l}{\cite{rei14}}\\
& $26\pm3$ &  &  & $218\pm6$ &  & $242_{-3}^{+10}$ &
\multicolumn{1}{l}{\cite{bov12}}\\
$7.7\pm0.9$ & $12.4\pm0.7$ &  &  & $236\pm3$ &  & $248.4$ &
\multicolumn{1}{l}{\cite{kaw19}}\\
& $12.1\pm7.6$ &  &  & $240\pm6$ &  & $252.1$ &
\multicolumn{1}{l}{\cite{hua16}}\\
$\approx11.1$ &  & $\approx7.8$ &  & $229\pm0.2$ & $\left(  -,-,-,\approx
2\%-5\%\right)  $ & $\approx245.8$ & \multicolumn{1}{l}{\cite{eil19}}\\
& $24\pm1$ &  &  &  & $\pm2\left(  V_{c}\right)  \pm5\left(  \text{large
scale}\right)  $ &  & \multicolumn{1}{l}{\cite{bov15}}\\
&  &  &  & $238\pm9$ &  & $250\pm9$ & \multicolumn{1}{l}{\cite{sch12}}\\
&  &  &  & $233.6\pm2.8$ &  &  & \multicolumn{1}{l}{\cite{mro19}}\\
&  &  & $\approx230$ &  &  &  & \multicolumn{1}{l}{\cite{ga1}}\\
&  &  &  & $224\pm13$ &  &  & \multicolumn{1}{l}{\cite{kop10}}\\
&  &  &  & $217\pm6$ &  &  & \multicolumn{1}{l}{\cite{weg19}}\\
&  &  &  &  &  & $247.4\pm1.4$ & \multicolumn{1}{l}{\cite{gra19}}\\
&  &  &  &  &  & $253\pm6$ & \multicolumn{1}{l}{\cite{hay18}}\\
\hline

\end{tabular}
%\end{center}$V
%\end{adjustwidth}
\end{table}

\subsection{Rotation Curves}
\label{velcur}From now, we will substitute galactic velocity $\mathbf{v(r)}$
in Equation~(\ref{r2a}) by $\mathbf{v(r)}$ $\rightarrow\mathbf{v(r)}+\Delta
\mathbf{V}_{\odot}$, which is the velocity related to the LSR:%
\vspace{-6pt}
\begin{align}
\mathbf{v(r)}  &  =\mathbf{V}_{G}\left(  \mathbf{R}\right)  +\Delta
\mathbf{v}\left(  \mathbf{R}\right)  -\mathbf{V}_{G}\left(  \mathbf{R}_{\odot
}\right)  ,\label{r10a}\\
\left\langle \mathbf{v(r)}\right\rangle  &  =\mathbf{V}_{G}\left(
\mathbf{R}_{\odot}\mathbf{+r}\right)  -\mathbf{V}_{G}\left(  \mathbf{R}%
_{\odot}\right)  . \label{r10c}%
\end{align}

\begin{figure}[h]%[H]
%\begin{adjustwidth}{-\extralength}{0cm}
\centering\includegraphics[width=170mm]{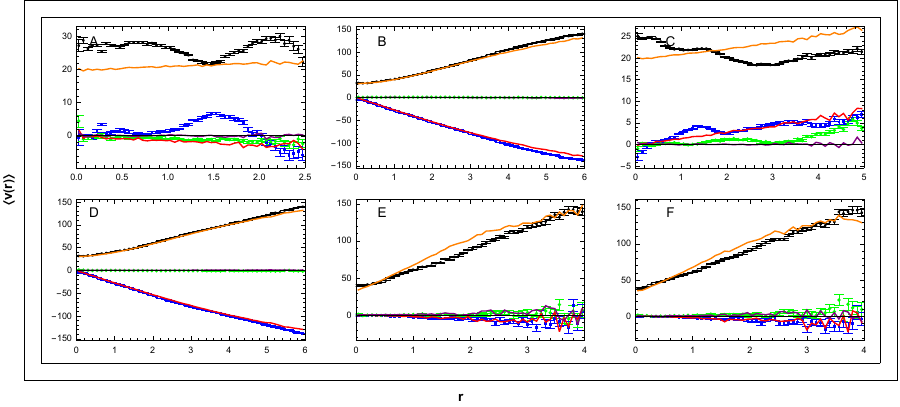}
%\end{adjustwidth}
\caption{Dependence of mean
velocity ($v_{l}$---blue, $v_{b}$---green, and $v_{gal}$---black) on distance $r$ in
the local rest frame at $\mathbf{R}_{\odot}$ in sectors (\textbf{A}--\textbf{F}).  For the corresponding curves ($v_{l}$---red, $v_{b}$---purple, and $v_{gal}$---orange) of the simulation model, see Sections~\ref{colmot} and  \ref{cosida}. Units: r[kpc], v[km/s].}%
\label{fg6}%
\end{figure}

\noindent In this frame, the input data in Equation~(\ref{me8}) are modified with the use of our
$\Delta\mathbf{V}_{\odot}$ as%
\vspace{-6pt}
\begin{align}
v_{l}  &  \rightarrow v_{l}+\Delta\mathbf{V}_{\odot}\mathbf{.n}_{l},\qquad
v_{b}\rightarrow v_{b}+\Delta\mathbf{V}_{\odot}\mathbf{.n}_{b},\label{r10b}\\
v_{gal}  &  \rightarrow\sqrt{\left(  v_{l}+\Delta\mathbf{V}_{\odot}%
\mathbf{.n}_{l}\right)  ^{2}+\left(  v_{b}+\Delta\mathbf{V}_{\odot}%
\mathbf{.n}_{b}\right)  ^{2}}. \label{r10d}%
\end{align}
After this substitution, Figure~\ref{fg4} is replaced by the top panel in
Figure~\ref{fg6}. The new $v_{l},v_{b},v_{gal}$ refer to the LSR whose velocity is%
\vspace{-6pt}
\begin{equation}
\mathbf{V}_{LSR}=\mathbf{V}_{G}\left(  \mathbf{R}_{\odot}\right)  =\left(
0,V_{0},0\right)  . \label{r11}%
\end{equation}

\begin{figure}[h]%[H]
%\begin{adjustwidth}{-\extralength}{0cm}
\centering\includegraphics[width=170mm]{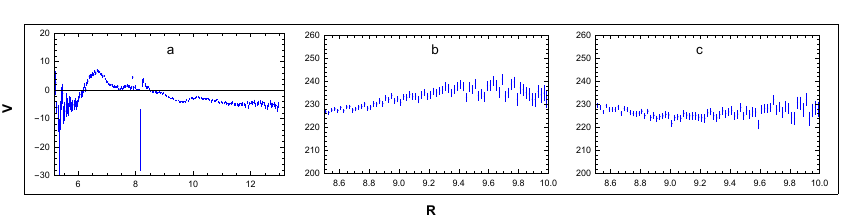}
%\end{adjustwidth}
\caption{Velocity curves
$V=V_{0}(R)-V_{0}(R_{\odot})$ in\ sectors A+C (panel (\textbf{a})) and $V=V_{0}(R)$ in
sectors B and D (panels (\textbf{b},\textbf{c})).
Units: R[kpc], v[km/s].}%
\label{fg7}%
\end{figure}

\noindent The combination of the new panels A and C, which represents the RC, is shown in
Figure~\ref{fg7}a.

\begin{figure}[h]%[H]%[t]
\includegraphics[width=115mm]{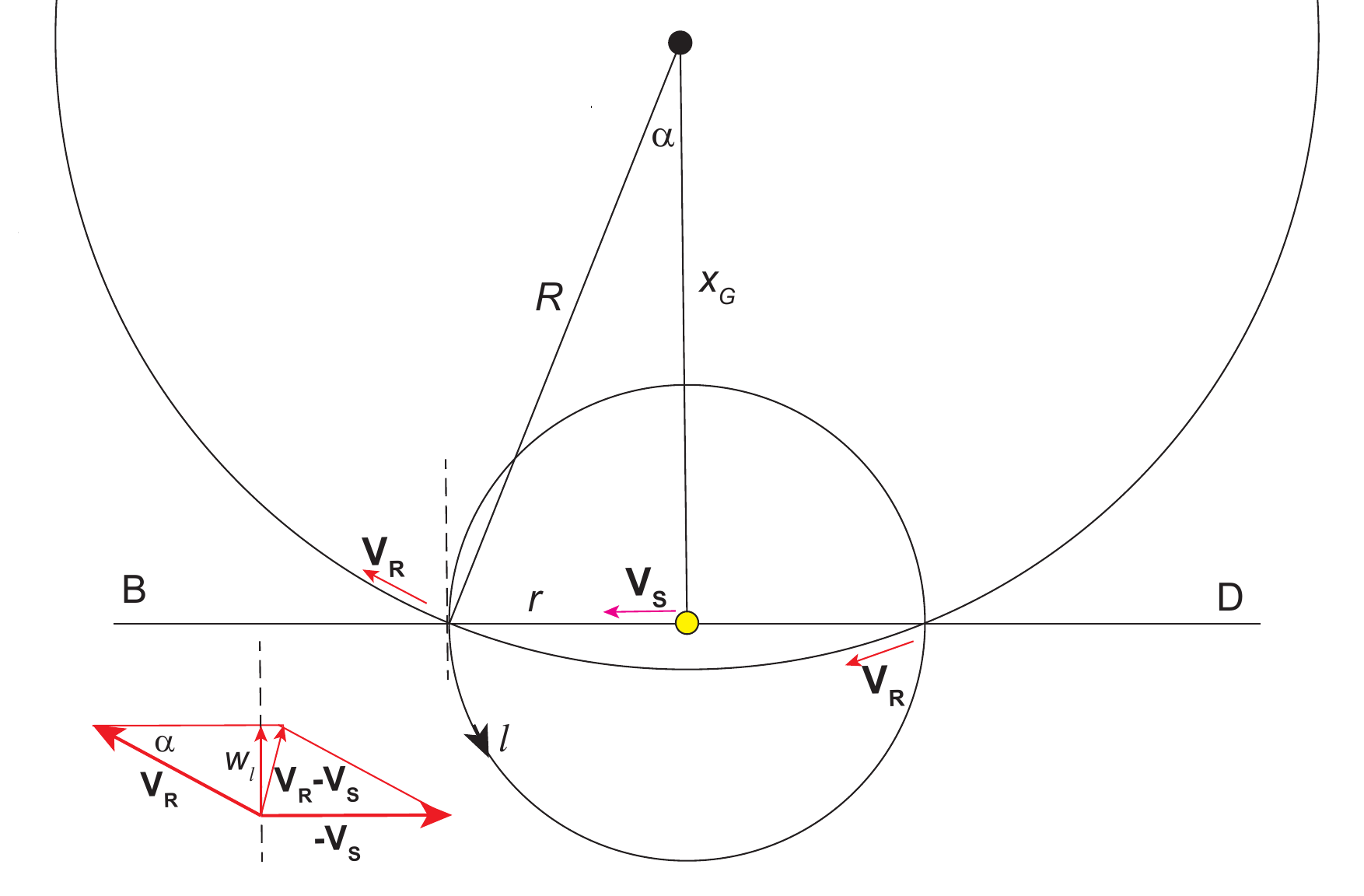}\caption{Rotation of MW as seen
in sectors B and D. Here the symbols \textbf{V}$_{\mathbf{R}}$ and
\textbf{V}$_{\mathbf{S}}$ stand for $\mathbf{V}_{G}\left(  \mathbf{R}\right)  $ and
$\mathbf{V}_{G}\left(  \mathbf{R}_{\odot}\right)  .$}%
\label{fg5}%
\end{figure}

Another representation of the RC is obtained from panels B and D. For
$\left\vert Z\right\vert \approx 0$ and $r>0,$ the term $\mathbf{w}=\mathbf{V}%
_{G}\left(  \mathbf{R}_{\odot}\mathbf{+r}\right)  -\mathbf{V}_{G}\left(
\mathbf{R}_{\odot}\right)  $ in Equation~(\ref{r10c}) and its transverse projection
$\left\langle w_{l}\right\rangle $ are calculated as suggested in
Figure~\ref{fg5} from two similar orthogonal triangles with angle $\alpha.$ We obtain
\begin{equation}
\left\langle w_{l}\right\rangle =\frac{r}{R}V_{0}\left(  R\right)  ;\qquad
R=\sqrt{r^{2}+R_{\odot}^{2}}. \label{r8}%
\end{equation}
Since $w_{l}=-v_{l}$, we obtain
\begin{align}
\left\langle v_{l}\right\rangle  &  =-\frac{r}{R}V_{0}\left(  R\right)
,\label{r9}\\
V_{0}\left(  R\right)   &  =-\frac{R}{r}\left\langle v_{l}\right\rangle .
\label{r10}%
\end{align}

\noindent The corresponding RCs are shown in panels b,c of Figure~\ref{fg7}. These curves are a different representation of the $\left\langle v_{l}\right\rangle $ curves in panels B and D of Figure~\ref{fg6}.

Relation (\ref{r9}) holds only for sectors B and D, where $l\approx\pm\pi/2$
and $Z\approx0$ (or $b\approx0$). In the general case, the geometry is more
complicated. With the use of (\ref{r2d}) and (\ref{r10c}), we have%
\vspace{-6pt}
\begin{equation}
\left\langle v_{l}\right\rangle =\left(  \mathbf{V}_{G}\left(  \mathbf{R}%
_{\odot}\mathbf{+r}\right)  -\mathbf{V}_{G}\left(  \mathbf{R}_{\odot}\right)
\right)  \mathbf{.n}_{l}. \label{r15}%
\end{equation}

\vspace{-6pt}
\vspace{-6pt}
\vspace{-12pt}

\noindent This relation can be modified as%

\vspace{-6pt}
\begin{equation}
\left\langle v_{l}\right\rangle =~-\left(  V_{G}\left(  \mathbf{R}_{\odot
}\mathbf{+r}\right)  \mathbf{N}_{\Phi}\left(  \mathbf{R}_{\odot}%
\mathbf{+r}\right)  -V_{0}\mathbf{N}_{\Phi}\left(  \mathbf{R}_{\odot}\right)
\right)  \mathbf{.n}_{l}. \label{r17}%
\end{equation}
With the use of relations (\ref{me3a}), (\ref{me1}), and (\ref{me2t}), vector  $\mathbf{N}_{\Phi}$
can be expressed in galactic coordinates $r,l,b$, and after a few further
modifications, we obtain%

\begin{figure}[h]%[H]
%\begin{adjustwidth}{-\extralength}{0cm}
\centering\includegraphics[width=170mm]{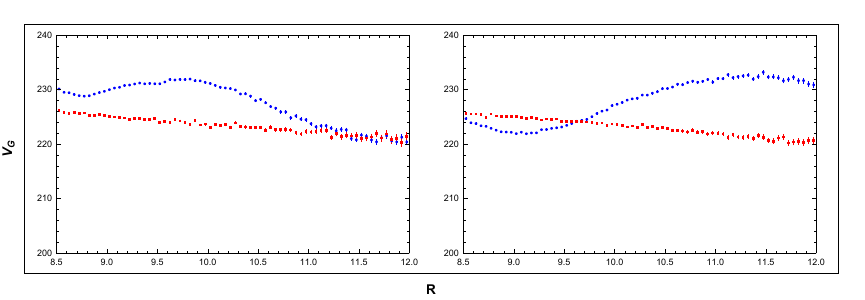}
%\end{adjustwidth}
\caption{RCs in sectors
Q$_{2}$ (\textbf{left}) and Q$_{4}$ (\textbf{right}): data (blue) and simulation (red).  For simulated curves, see Sections~\ref{colmot} and  \ref{cosida}. Units:
R[kpc], $V_{G}$[km/s]. }%
\label{fg8}%
\end{figure}

\begin{figure}[h]%[H]
%\begin{adjustwidth}{-\extralength}{0cm}
\centering\includegraphics[width=170mm]{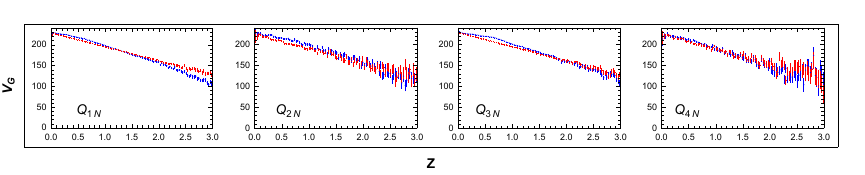}
\includegraphics[width=170mm]{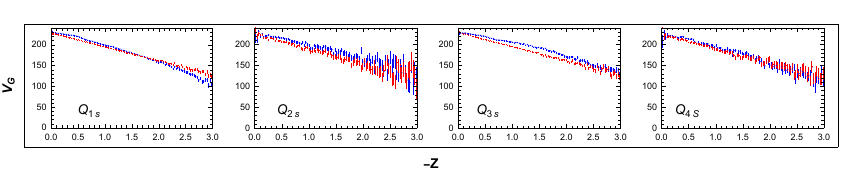}
%\end{adjustwidth}
\caption{Velocity curves
$V_{G}\left(  \left\vert Z\right\vert \right)  $ in sectors Q$_{1S}-$Q$_{4S}$
and Q$_{1N}-$Q$_{4N}$: data (blue) and simulation (red). For simulated curves, see Sections~\ref{colmot} and  \ref{cosida}. Units: Z[kpc],
$V_{G}$[km/s]. }%
\label{fg8NS}%
\end{figure}

\vspace{-6pt}
\begin{equation}
\left\langle v_{l}\right\rangle =\gamma V_{G}\left(  \mathbf{R}_{\odot
}\mathbf{+r}\right)  -V_{0}\cos l \label{r18}%
\end{equation}
and if $\gamma\neq0$, then
\vspace{-6pt}
\begin{equation}
V_{G}\left(  \mathbf{R}_{\odot}\mathbf{+r}\right)  =\ \frac{\left\langle
v_{l}\right\rangle +V_{0}\cos l}{\gamma};\qquad\gamma=\frac{R_{\odot}\cos
l-r\cos b}{\sqrt{R_{\odot}^{2}+r^{2}\cos^{2}b-2R_{\odot}r\cos l\cos b}%
}.\label{r18a}%
\end{equation}
One can check that in sectors B and D, where $b\approx 0$,\ \ $l\approx \pm \pi/2$ and where we have assumed $V_{G}\approx
V_{0}\left(  R\right)  $, relation (\ref{r18}) reduces to (\ref{r9}). Relation
(\ref{r18a}) allows us to analyse $V_{G}\left(  \mathbf{R}\right)  $ not only
in narrow sectors B and D but also in the wider regions, which can provide
higher statistics with smaller errors. This relation is not suitable for the
reconstruction of $V_{G}$ in the vicinity of singularity $\gamma\approx0$
\ $\left(  R_{\odot}\cos l\approx r\cos b\right)  $. In Figure~\ref{fg8}
(blue points), we show RCs obtained with the use of Equation~(\ref{r18a}) in
the sectors Q$_{2}$ and Q$_{4}$. In the analysed area, we observe irregular
fluctuations in the rotation velocity: $\Delta V_{G}/V_{G}\approx5\%$. The
relation allows us to calculate rotation velocity not only in the galactic
plane but also outside the plane. In Figure~\ref{fg8NS}, we show the velocity curves $V_{G}\left(  \left\vert Z\right\vert
\right)  $ calculated in sectors Q$_{1S}-$Q$_{4S}$ and Q$_{1N}-$Q$_{4N}$.

%\vspace{-12pt}
%\vspace{-6pt}
%\vspace{-6pt}
%\vspace{-9pt}

\subsection{Six Parameters of the MW Collective Rotation}
\label{colmot}
\begin{figure}[h]%[H]
%\begin{adjustwidth}{-\extralength}{0cm}
\centering\includegraphics[width=165mm]{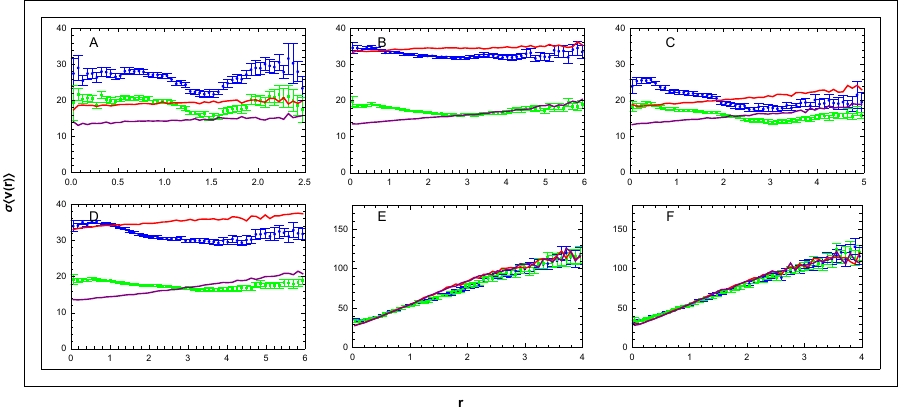}
%\end{adjustwidth}
\caption{Dependence of the
standard deviations of the mean velocity ($v_{l}$---blue and $v_{b}$---green) on
the distance $r$ in sectors (\textbf{A}--\textbf{F}). For simulated curves ($v_{l}$---red and $v_{b}$---purple), see Section~\ref{cosida}. Units: r[kpc], $\sigma$[km/s]. }%
\label{fg9}%
\end{figure}

Panels E and F at the top of Figure~\ref{fg6} provide further
important information. We observe $\left\langle v_{l}\right\rangle \approx0$
and $\left\langle v_{b}\right\rangle \approx0$, as expected in both narrow
cones pointing perpendicularly from the galactic plane, where positive and
negative $v_{l},v_{b}$ are equally abundant. On the other hand, the value
$\left\langle v_{gal}\right\rangle $ increases with distance from the plane.
This increase occurs in the galactic reference frame reflecting the
slowing of collective rotation in the Galactocentric frame, which increases with $r\approx\left\vert Z\right\vert $. The same
effect is seen even more clearly in Figure~\ref{fg8NS}. Important information is
obtained from Figure~\ref{fg9}, where dependencies of standard deviations are shown. The
increasing standard deviations in panels E and F suggest a less collective
but more disorderly motion of high velocities away from the galactic plane.

In the distribution (\ref{ge3}), we assume in a\textit{ }first approximation that
\vspace{-6pt}
\begin{equation}
\sigma_{\alpha}=\sigma_{\alpha0}+\sigma_{\alpha1}\left\vert Z\right\vert
;\qquad\alpha=R,\Theta,\Phi, \label{r12}%
\end{equation}
where $\sigma_{\alpha0}$ and $\sigma_{\alpha1}$ are the free parameters. We
proceed as follows in their setting:

(i) %MDPI: Please confirm if the italics is unnecessary and can be removed. The following highlights are the same. PZ: ok, italics removed..
From the data panels A--D in Figure~\ref{fg9} where $Z\approx0$, we estimate the first approximation:%
\vspace{-6pt}
\begin{align}
\sigma_{\Theta0}  &  \approx\left\langle \sigma_{b}\right\rangle
\qquad\text{in sectors A-D,}\label{r12a}\\
\sigma_{R0}  &  \approx\left\langle \sigma_{l}\right\rangle \qquad\text{in
sectors B,D and at small }r/R_{\odot}\text{,}\label{r12b}\\
\sigma_{\Phi0}  &  \approx\left\langle \sigma_{l}\right\rangle \qquad\text{in
sectors A,C.} \label{12d}%
\end{align}
In directions other than A--D, the relations between $\sigma$s in the galactic and Galactocentric frames are more complex. The final tuning of these three parameters is performed in Section~\ref{cosida} using further curves.

(ii) The data panels E and F (where
$r\approx\left\vert Z\right\vert $) show in the region of the peaks
(\mbox{$r\lesssim2.5$ kpc}, see Figure~\ref{fg3}, panels E,F) a linear increase in the corresponding mixture of $\sigma$s with $\left\vert Z\right\vert $.
To obtain the parameters $\sigma_{\alpha1}$, we analysed the
respective distributions in all the sectors listed in Table~\ref{Tab1}. We found
that the optimal shape of the distribution $V_{\Phi}$ in (\ref{ge3})\ is
asymmetric, having different $\sigma_{\Phi}^{\pm}$ for the two opposite
orientations, where $+(-)$ means \textit{in (against)} the rotation direction:%
\vspace{-6pt}
\begin{equation}
\sigma_{\Phi}^{+}=\sigma_{\Phi0},\qquad\sigma_{\Phi}^{-}=\sigma_{\Phi0}%
+\sigma_{\Phi1}^{-}\left\vert Z\right\vert .\label{r13c}%
\end{equation}
So only $\sigma_{\Phi}^{-}$ depends on $\left\vert Z\right\vert $ and
$\sigma_{\Phi}^{+}$ does not. This
asymmetry also reflects the effective deceleration $\left\langle \Delta
V_{\Phi}\right\rangle $ of the collective rotation for larger $\left\vert
Z\right\vert $ as mentioned above. We have
\begin{align}
\left\langle \Delta V_{\Phi}\right\rangle =\frac{1}{N}\left(  \int_{-\infty
}^{0}V_{\Phi}\exp\left(  -\frac{1}{2}\left(  \frac{V_{\Phi}}{\sigma_{\Phi}%
^{-}}\right)  ^{2}\right)  dV_{\Phi}+\int_{0}^{\infty}V_{\Phi}\exp\left(
-\frac{1}{2}\left(  \frac{V_{\Phi}}{\sigma_{\Phi}^{+}}\right)  ^{2}\right)
dV_{\Phi}\right)  ;\label{r13a}\\
N=\int_{-\infty}^{0}\exp\left(  -\frac{1}{2}\left(  \frac{V_{\Phi}}%
{\sigma_{\Phi}^{-}}\right)  ^{2}\right)  dV_{\Phi}+\int_{0}^{\infty}%
\exp\left(  -\frac{1}{2}\left(  \frac{V_{\Phi}}{\sigma_{\Phi}^{+}}\right)
^{2}\right)  dV_{\Phi}.\nonumber
\end{align}
After integration we obtain
\vspace{-6pt}
\begin{equation}
\left\langle \Delta V_{\Phi}\right\rangle =\sqrt{\frac{2}{\pi}}\left(
\sigma_{\Phi}^{+}-\sigma_{\Phi}^{-}\right)  =-\sqrt{\frac{2}{\pi}}\sigma
_{\Phi1}^{-}\left\vert Z\right\vert .\label{r13b}%
\end{equation}
The important parameter $\sigma_{\Phi1}^{-}$\ is obtained by the fit from
Figure~\ref{fg8NS}, which suggests dependence:
\vspace{-6pt}
\begin{align}
V_{G}\left(  \left\vert Z\right\vert \right)  =V_{0}+\left\langle \Delta
V_{\Phi}\right\rangle \approx V_{0}-\varkappa\left\vert Z\right\vert
;\qquad\varkappa=\sqrt{\frac{2}{\pi}}\sigma_{\Phi1}^{-}\doteq
33.5\mathrm{km\,s}^{-1}\mathrm{kpc}^{-1}.\label{r13d}%
\end{align}

\begin{figure}[h]%[H]
%\begin{adjustwidth}{-\extralength}{0cm}
\centering\includegraphics[width=155mm]{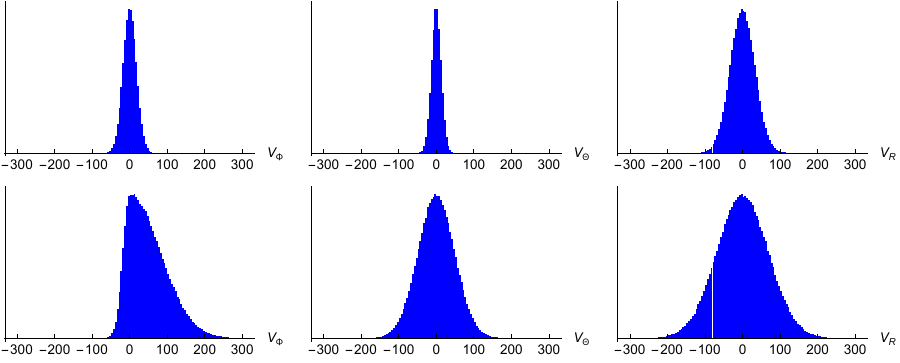}
%\end{adjustwidth}
\caption{Distribution of
simulated velocities $V_{\Phi},V_{\Theta}$, and $V_{R}$ at $Z=0$ (\textbf{upper row})
and $Z=1.5$ kpc (\textbf{lower row}). Units: $V$[km/s]. }%
\label{fg18}%
\end{figure}

For the remaining two parameters, the analysis showed that $\sigma_{\Theta1}$
$\approx\sigma_{R1}$ is a good approximation. We denote $\sigma_{1}$
$\equiv\sigma_{R1}=\sigma_{\Theta1}$: this last free parameter was set up from
the tuning of the slope in simulation panels E and F in Figure~\ref{fg9}. A list
of the six resulting parameters controlling simulation (\ref{ge3}) is given in
Table~\ref{Tab3}. The relation (\ref{r13d}) shows that the MW rotation at
$Z_{\odot}\approx0.0208$ kpc is $0.7$ km/s lower than at $Z=0$, which is
significantly less than the total error in determining $V_{0}$.
Therefore, we neglected $Z_{\odot}$ in our calculation. The Wolfram
Mathematica code of the generator is available on the website
\url{https://www.fzu.cz/~piska/Catalogue/genkinJAN25.nb} {(accessed on 20 January 2025).} %MDPI: Please add the access date (format: Date Month Year), e.g., accessed on 1 January 2020. PZ: ok, date inserted...
Figure~\ref{fg18} shows the examples of the distribution of simulated velocities
$V_{\Phi},V_{\Theta}$, and $V_{R}$. The simulated distributions $V_{\Theta}$
and $V_{R}$ are symmetric for any $\left\vert Z\right\vert $, and distribution
$V_{\Phi}$ is asymmetric for $\left\vert Z\right\vert >0$.

\vspace{-6pt}
\vspace{-12pt}
\begin{table}[h]%[H]%[ptb]
\caption{Monte Carlo %MDPI: Please move Table 3 after where it was first mentioned and keep ref citation order correct..PZ: order is ok, please do not change...%MDPI: We moved Figure 12 together with Table 3 to its current place, after the discussions of them, please check and confirm the place.
simulation model parameters and corresponding parameters from
other analyses. Velocity $V_{0}$ is taken from Table~\ref{Tab2}. }%
\label{Tab3}%
%\begin{center}%
%\begin{adjustwidth}{-\extralength}{0cm}
\centering
\begin{tabular}{ccccccc}%
\hline
\boldmath\textbf{$\sigma_{\Theta0}$~[km~s$^{-1}$]} & \boldmath\textbf{$\sigma_{\Phi0}$~[km~s$^{-1}$]} & \boldmath\textbf{$\sigma
_{R0}$~[km~s$^{-1}$]} & \boldmath\textbf{$\sigma_{1}$~[km~s$^{-1}$kpc$^{-1}$]} & \boldmath\textbf{$\sigma_{\Phi1}%
^{-}$~[km~s$^{-1}$kpc$^{-1}$]}& \boldmath\textbf{$V_{0}$~[km~s$^{-1}$]} & \textbf{Ref.}\\\hline
$13$ & $18$ & $33$ & $24$ & $42$ & $228$ & \multicolumn{1}{l}{this work}\\
$18.03\pm0.03$ & $24.35\pm0.04$ & $36.81\pm0.07$ & x & x & x &
\multicolumn{1}{l}{\cite{ang20}}\\
$11$ & $20$ & $31$ & x & x & x & \multicolumn{1}{l}{\cite{vie22}}\\
\hline
\end{tabular}
%\end{center}
%\end{adjustwidth}
\end{table}
\vspace{-12pt}

\subsection{Comparison of Simulation Model with Data}
\label{cosida}
The comparison is performed as follows:

(1) The position $\left(  r,l,b\right)  $ of the source in the galactic reference
frame is with the use of Equation~(\ref{me1}) transformed to the Galactocentric
spherical frame $\left(  R,\Phi,\Theta\right)  $. For this position, the
velocity $\mathbf{V}^{gen}\left(  \mathbf{R}\right)  $ defined by (\ref{ge1})
is generated according to distribution (\ref{ge3}) with the parameters from
Table~\ref{Tab3}.

(2) Using (\ref{r11})\ and (\ref{me3b})\, we obtain the corresponding LSR
transverse velocities:%
\vspace{-6pt}
\begin{equation}
v_{\alpha}^{gen}\mathbf{(r)}=\left(  \mathbf{V}^{gen}\left(  \mathbf{R}%
\right)  -\mathbf{V}_{LSR}\right)  \mathbf{.n}_{\alpha};\qquad\alpha=l,b.
\label{r14}%
\end{equation}
So for any source defined by input $\left(  r,l,b,v_{l},v_{b}\right),  $ we
generate a vector $\left(  r,l,b,v_{l}^{gen},v_{b}^{gen}\right)  $ and then
create the desired distributions from both.

(3) These distributions obtained from the input data and simulations will now
be compared. The comparison in Figure~\ref{fg6} shows overall very good agreement between data and
simulation for panels B,D,E,F. In panels A and C, the simulation model does not reproduce local kinematic substructures generating deviations $\Delta
v\lesssim10$ km/s. The presence of these substructures shows the precision
with which we work. At the same time, the fluctuations are not noticeable in
the other panels because their velocity scale is coarser.

The simulation in panel A(C) indicates that the velocity $\left\langle
v_{l}(r)\right\rangle _{S}$ decreases (increases) with $r$, despite the constant parameter
$V_{0}$. This small effect is because we are working inside the angle $b=\pm5$
deg, which means a slight linear increase in average $\left\vert Z\right\vert
$ and correspondingly some deceleration with $r$. So, a corresponding
correction would therefore be necessary to evaluate the RC in this sector more
accurately. We checked that for smaller angles $\left\vert b\right\vert $,
this effect disappears.

Figure~\ref{fg9} shows the corresponding velocity dispersion dependencies. In the
upper panels A--D, we again observe fluctuations that are not present in
the simulation. In panels E,F with a coarser scale, the fluctuations are not
noticeable, and agreement with the simulation is very good. Averaged data in panels B-D do not contradict the simulation. The difference between data and simulation in panel A is more obvious and will be analysed~below.

Relatively small velocity fluctuations ($\Delta V_{G}\lesssim10$ km/s, $\Delta V_{G}/V_{G}\lesssim0.05$) also
appear in the RC in Figure~\ref{fg8}. The slightly decreasing simulated RC is due
to the shape of the sectors Q$_{2}$ and Q$_{4}$, where a larger $R$ correlates
with a larger average $\left\vert Z\right\vert $, implying a smaller $V_{G}$.
Figure~\ref{fg8NS} shows that the simulation of decreasing $V_{G}\left(
\left\vert Z\right\vert \right)  $ controlled by the fitted parameter
$\sigma_{\Phi1}^{-}$ in Equation~(\ref{r13d}) agrees well with the data.

\begin{figure}[h]%[H]
%\begin{adjustwidth}{-\extralength}{0cm}
\centering\includegraphics[width=170mm]{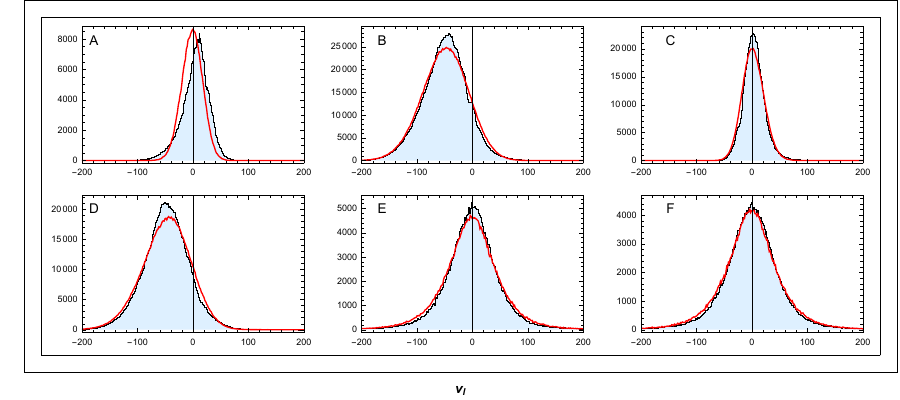}\\
\includegraphics[width=170mm]{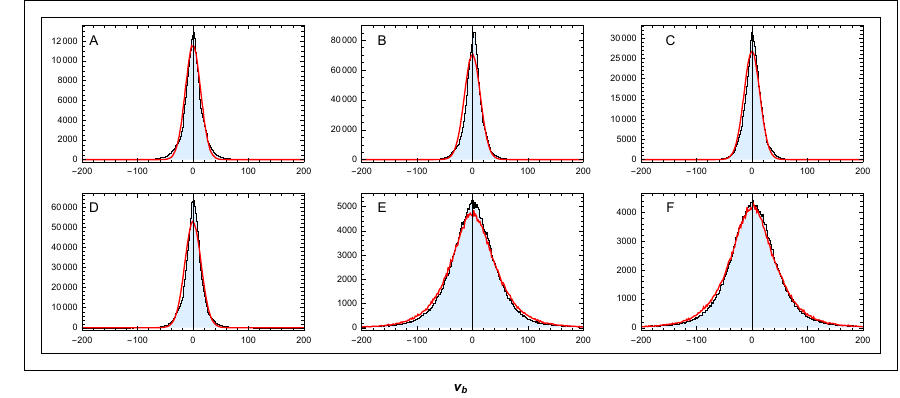}
%\end{adjustwidth}
\caption{{Distributions} %MDPI: Please use commas to separate thousands for numbers with five or more digits (not four digits) in the picture, e.g., "10000" should be "10,000". PZ: pictures are produced in this format...
of $v_{l}$ and $v_{b}$ in sectors (\textbf{A}--\textbf{F}): data (light blue) and simulation model (red). Unit: v[km/s].
Binning: 1.6 km/s.}%
\label{fg10}%
\end{figure}

The good agreement of the simulations with the data is confirmed by other
results. Figure~\ref{fg10} shows distributions of $v_{l}$ and $v_{b}$ in
sectors A--F along with the corresponding distributions obtained from
simulations. With the help of this data, the final tuning of the parameters $\sigma_{\Theta0}$ and $\sigma_{\Phi0}$ was achieved. An apparent disagreement with the simulation occurs in sector A (where $R<R_{\odot}$).  It may be a manifestation of the asymmetric drift effect. In the opposite sector C
(where $R>R_{\odot}$), the asymmetry is not manifested. The distributions of
$v_{l}$ in both sectors A and C are copies (up to a constant shift) of the
distribution of orbital velocities in the Galactocentric frame. Note the shift in sectors B
and D resulting from the decrease in $\left\langle v_{l}\right\rangle $ in  panels B and D in Figure~\ref{fg6}. As expected, there is a clear symmetry between the two panels.
%\vspace{-6pt}

\begin{figure}[h]%[H]
%\begin{adjustwidth}{-\extralength}{0cm}
\centering\includegraphics[width=170mm]{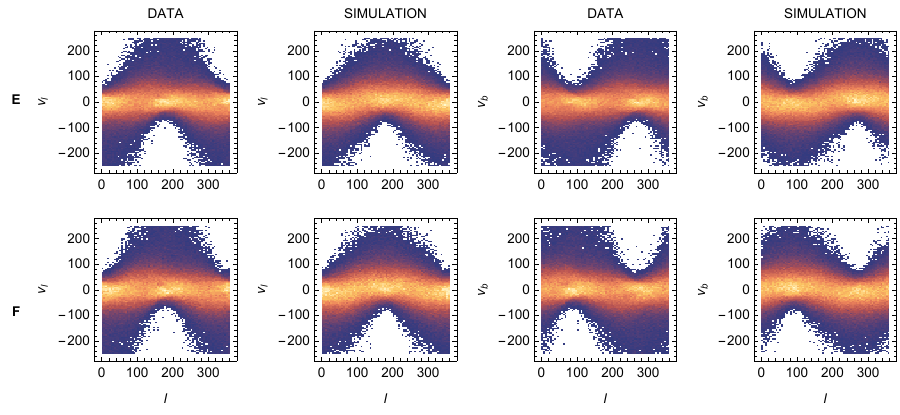}
%\end{adjustwidth}
\caption{Histograms
$\left(  v_{l},l\right)  $ and $\left(  v_{b},l\right)  $ in sectors E and F:
data and simulation. Units: $l$[deg], $v$[km/s]. Binning $l,v$:
$5\ $deg$,\ 5$~km/s.}%
\label{fg11}%
\end{figure}

\begin{figure}[t]
%\begin{figure}[H]
%[t]
\includegraphics[width=120mm]{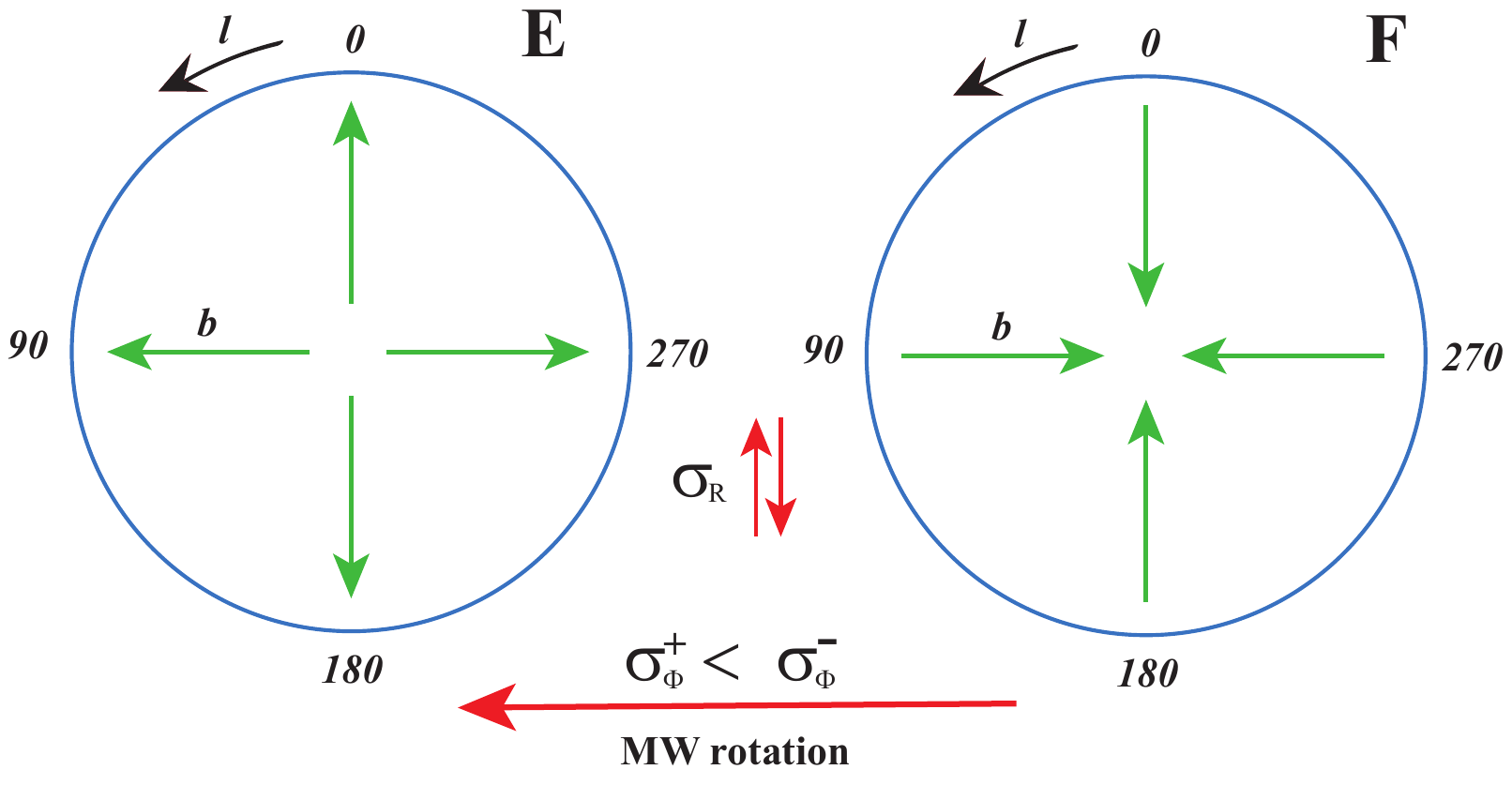}\caption{Asymmetry
$\sigma_{\Phi}^{+}<\sigma_{\Phi}^{-}$ in the Galactocentric reference frame
generates asymmetries in the galactic frame, see text and Figure~\ref{fg11}.}%
\label{fg19}%
\end{figure}

The important result is shown in Figure~\ref{fg11}. The asymmetry of histograms
$\left(  v_{l},l\right)  $ and $\left(  v_{b},l\right)  $ in sectors E and F
with the white areas reflects different projections of the asymmetry represented by Equation (\ref{r13c}). The yellow--orange region at $v_{l},v_{b}=0$ corresponds to the
peaks in Figure~\ref{fg10}E,F,  which are the integrals of the histograms over $l$.
The shape of
histograms can be explained using Figure~\ref{fg19}.
Distributions of $v_{l}$ for $l\approx0^{\circ},90^{\circ},180^{\circ
},270^{\circ}$ are controlled by the parameters $\sigma_{l}^{\pm}$. Their connection with $\sigma_{\Phi}^{\pm}$ and $\sigma_{R}$
can be deduced from the figure. At $l\approx0^{\circ},$ the
directions of $v_{l}$ and the MW rotation are identical, so $\sigma_{l}^{\pm
}\approx\sigma_{\Phi}^{\pm}$. But at $l=180^{\circ}$, the two directions are
opposite, so $\sigma_{l}^{\pm}\approx\sigma_{\Phi}^{\mp}$. At $l\approx
90^{\circ}(270^{\circ})$, the situation is a little more complicated. If  $\left\vert Z\right\vert/R $ or $\Theta$ are small (which is almost
our case, see panels E,F in Figure~\ref{fg3}, where $r$ $\approx\left\vert Z\right\vert
$), then the $v_{l}$ direction can be approximated by vector $\mathbf{R}$, so $\sigma_{l}^{\pm}\approx
\sigma_{R}$, similarly for $v_{b}$ distributions
controlled by the parameters $\sigma_{b}^{\pm}$, which are also related to
$\sigma_{\Phi}^{\pm}$ and $\sigma_{R}$. Also here, the agreement between the data and the model is very good.  Clear evidence for our explanation was provided by the MC simulation: after setting the symmetric $\sigma_{\Phi}^{+}=\sigma_{\Phi}^{-}$, the asymmetry in the figure disappears.

\vspace{-6pt}
\vspace{-6pt}
\vspace{-6pt}
\begin{figure}[h]%[H]
%\begin{adjustwidth}{-\extralength}{0cm}
\centering\includegraphics[width=170mm]{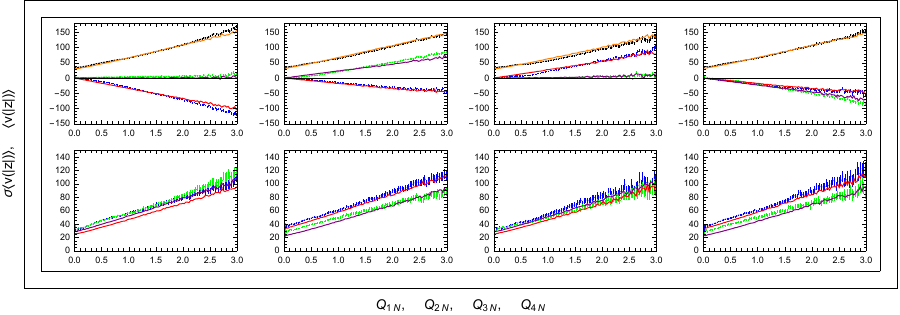}\\
\

\includegraphics[width=170mm]{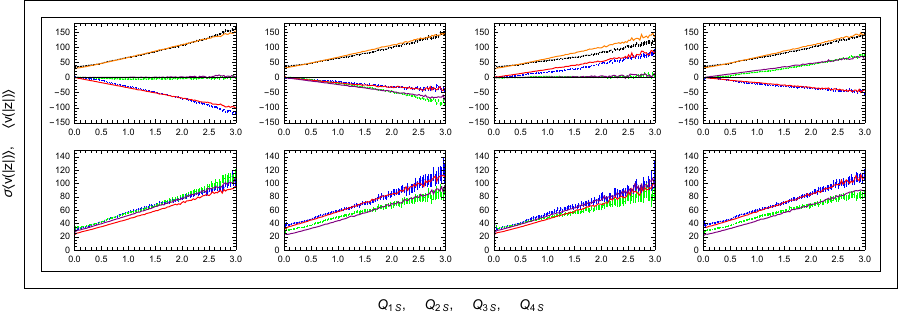}
%\end{adjustwidth}
\caption{Dependence of mean
velocity and its dispersion ($v_{l}$---blue, $v_{b}$---green, and $v_{gal}$---black)
on distance $|z|$ in sectors Q$_{1N}$--Q$_{4N}$ and Q$_{1S}$--Q$_{4S}$: data and
simulation model ($v_{l}$---red, $v_{b}$---purple, and $v_{gal}$---orange). Units: z[kpc], v[km/s].}%
\label{nf15}%
\end{figure}

\begin{figure}[h]%[H]
%[ptb]
%%[t]
%\begin{adjustwidth}{-\extralength}{0cm}
\centering\includegraphics[width=170mm]{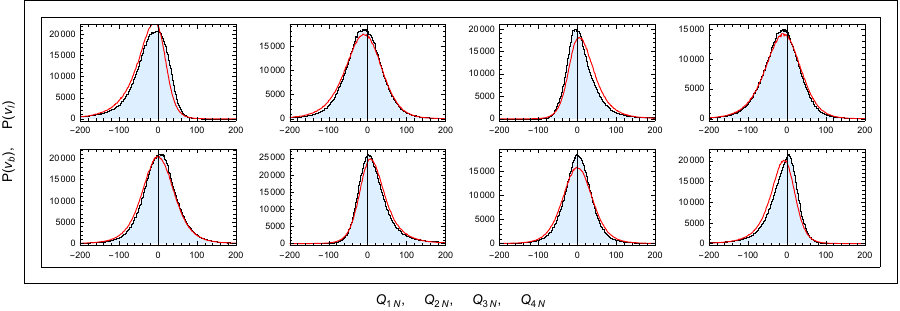}\\
\

\includegraphics[width=170mm]{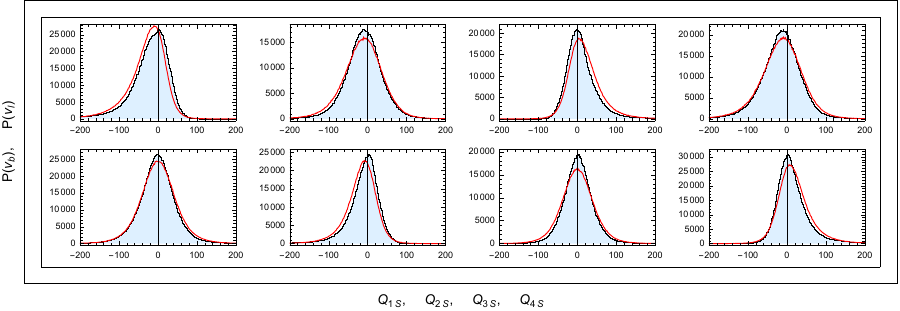}
%\end{adjustwidth}
\caption{{Distributions} %MDPI: Please use commas to separate thousands for numbers with five or more digits (not four digits) in the picture, e.g., "10000" should be "10,000". PZ: pictures are produced in this format...
of $v_{l}$ and
$v_{b}$ in sectors Q$_{1S}-$Q$_{4S}$ and Q$_{1N}-$Q$_{4N}$: data (light blue) and simulation model (red). Unit: v[km/s]. Binning: 1.6 km/s. }%
\label{fg12}%
\end{figure}

\clearpage
Correct Monte Carlo parameter settings can be verified in wide Q-sectors outside the area
of the galactic plane. Figure~\ref{nf15} shows the $z-$dependence of mean values and
dispersions of velocity distributions in these sectors. The curves together
with the corresponding overall distributions of velocities in broad sectors
Q$_{1N}-$Q$_{4N}$ and Q$_{1S}-$Q$_{4S}$ in Figure~\ref{fg12} again confirm the very good agreement of the simulation with data. Note in particular the projections
$v_{l}$ in sectors Q$_{1N}$,Q$_{3N}$,Q$_{1S}$, and Q$_{3S}$ and $v_{b}$ in sectors
Q$_{2N}$,Q$_{4N}$,Q$_{2S}$, and Q$_{4S}$. This is also due to the asymmetry
expressed in Equation~(\ref{r13c}) that occurs for $\left\vert Z\right\vert >0$, as illustrated by the simulated $V_{\Phi}$ distribution in Figure~\ref{fg18}. Note also the expected symmetry between the corresponding Q$_{N}$ and Q$_{S}$ panels in Figure~\ref{fg12}.

%The good agreement of the simulations with the data is confirmed by other
%results.

\vspace{-6pt}


\section{Discussion and Conclusions}
\label{discussion}
(1) The results on local Sun's velocity $\boldsymbol{v}_{\odot}=\left(
U_{\odot},V_{\odot},W_{\odot}\right)  $, the MW circular velocity $V_{c}$\, and
average orbital velocity $V_{0}$ can be compared with the other measurements
presented in Table~\ref{Tab2}. The excellent agreement with others is obtained
for the rotational velocity of the Sun $V_{c}+V_{\odot}=V_{0}+\Delta
V_{y\odot}$. Within the measurement errors, this result perfectly agrees with
all the others, including a very accurate measurement \cite{gra19}. There is
also perfect agreement with the others for $W_{\odot}$ and $V_{0}$ and a good
agreement for $U_{\odot}$. However, differences in $V_{c}$ and $V_{\odot}$
from different measurements are larger, so our values agree with only some of
them (within errors).

\begin{figure}[h]%[H]
%\begin{adjustwidth}{-\extralength}{0cm}
\centering\includegraphics[width=170mm]{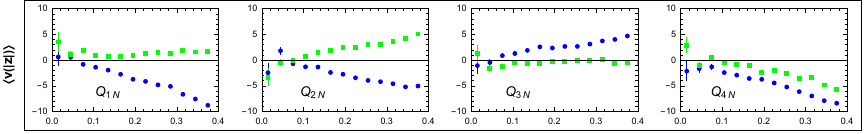}
\includegraphics[width=170mm]{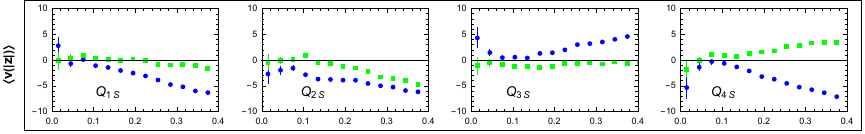}
\includegraphics[width=170mm]{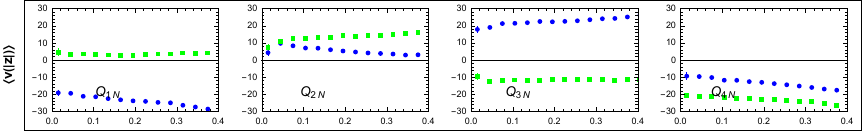}
\includegraphics[width=170mm]{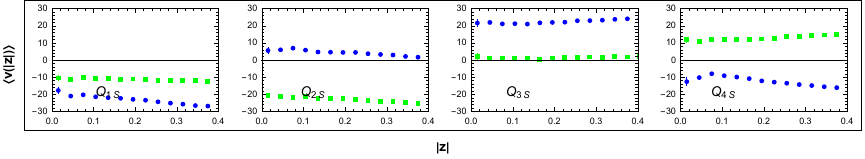}
%\end{adjustwidth}
\caption{Dependence of mean
velocity ($v_{l}$---blue and $v_{b}$---green) on distance $|z|$ from the galactic plane in sectors Q$_{1N}%
$--Q$_{4N}$ and Q$_{1S}$--Q$_{4S}$ in the LSR reference frame (\textbf{upper part}) and
the Sun's rest frame (\textbf{lower part}). Units: z[kpc], v[km/s].}%
\label{nf17}%
\end{figure}

Let us add a few remarks on our measurement of $\Delta\mathbf{V}_{\odot}$ and
$V_{0}$. Both velocities are obtained independently with the use of a direct
and model-independent method. The $\Delta\mathbf{V}_{y\odot}$ is measured equally as
$\Delta\mathbf{V}_{z\odot}$, see Figure~\ref{fg4VW}. For $r_{max}>0.5$ kpc, we obtain a (nearly)
constant \ in the band of small statistical errors. The calculation is based on data from different sectors, so the resulting systematic errors in Table~\ref{Tab2} are much larger than the statistical errors. The source of the systematic errors is mainly due to kinematic substructures outside the axial symmetry, which may vary from sector to sector.
The correct determination of $\Delta\mathbf{V}_{\odot}$
can be verified as follows. At the top half of Figure~\ref{nf17}, we show the dependencies of $\left\langle v(z)\right\rangle $
(magnified parts of Figure~\ref{nf15}) in the eight Q$_{S,N}$ sectors in the LSR
reference frame. This frame defines the $\Delta\mathbf{V}_{\odot}$ determined
from the data in sectors A-D, see Table~\ref{Tab2}. In these figures, we observe
that for $z\rightarrow0$, we have $\left\langle v_{l}(z)\right\rangle
,\left\langle v_{b}(z)\right\rangle $ $\rightarrow0$, which means that the
solar velocity related to the sectors A--D is the same as the velocity related
to the nearby stars in all sectors Q$_{S,N}$. This agreement is a simple test
that our velocity $\Delta\mathbf{V}_{\odot}$ (Equation~(\ref{cor5})) is correct. In the bottom
half of the figure, we have the same curves for comparison but in a galactic
reference frame (the Sun's rest frame). Note the different scales on the upper and lower parts.
%\vspace{-6pt}

(2) The determination of RC is based on the model-independent definition
(\ref{ge2}). In Figures~\ref{fg7} and \ref{fg8}, we show RCs measured in different
sectors of galactic longitudes. The curves are obtained with very high
precision, so as a result, we observe local fluctuations ($\lesssim10$ km/s)
in the structure of the MW rotation. These fluctuations correspond to the velocity
substructures and non-axisymmetric kinematic signatures mentioned in
Section~\ref{intro}. The fluctuations do not allow us to analyse the RC slope in
our limited range of $R$. Further, we showed that the collective rotation
velocity decreases for increasing $\left\vert Z\right\vert $, see
Figure~\ref{fg8NS}. A similar observation was reported in \cite{wan22,ang20}. The
slope of the curves is defined in \mbox{Equation~(\ref{r13d})} and can be compared with a
prediction \cite{bie18}{ (Figure 5)} %MDPI: please check if this figure belongs to ref 9. PZ: Yes, it belongs to this ref.
of the Besan\c{c}on model. Within the range
of our analysis (0--3~kpc), the average slopes agree very well, although the shape
of the curves is slightly different.

(3) Except for observed local fluctuations, the analysed kinematical
distributions are very well described by a minimal MW axisymmetric model based
on six free parameters in the Galactocentric reference frame. The model
describes a simplified scenario in which local velocity fluctuations are
averaged. The scale of averaged fluctuations increases with distance from
the galactic plane and is defined by the parameters of the
model in Table~\ref{Tab3}. The fluctuations are most
significant in the $\mathbf{N}_{R}$ direction, less in the $\mathbf{N_{\Phi}}$
direction, and least in the $\mathbf{N_{\Theta}}$ direction.

The analysis and simulation of kinematical distributions in the studied region
that are a part of the MW parameters also need other parameters related to our
laboratory: % EE: Please check that intended meaning has been retained.
its local 3D velocity $\Delta\mathbf{V}_{\odot}$, distance from
the galactic centre $R_{\odot}$, and its position $z_{\odot}$ above the
galactic plane (neglected). Thus, except for $R_{\odot}$, all the remaining
parameters that we obtained in the present analysis are listed in
Tables~\ref{Tab2} and \ref{Tab3}. For now, we ignore the slope of the RC (Equation~(\ref{r7b})), which has
in our region a very small effect \cite{eil19,rei14}. The model describes the MW rotation as follows.

(a) The rotation is strongly collective in the galactic disk plane with
relatively small Gaussian velocity fluctuations $\sigma_{\Phi0},\sigma
_{\Theta0},\sigma_{R0}$ around the much greater velocity $V_{0}$. This can be
seen in Figure~\ref{fg10} in panels $v_{l}$ for sectors A and C and panels
$v_{b}$ for sectors A--D. Wider and shifted distributions
$v_{l}$ for sectors B and D are due to the rotation effect shown in
Figure~\ref{fg5} and expressed in Equation~(\ref{r9}). Our first three parameters are
compared with corresponding galactic thin disc parameters obtained in another
study, see Table~\ref{Tab3}. The agreement with \cite{vie22}
is very good.

(b) The parameters $\sigma_{1}$ and $\sigma_{\Phi1}^{-}$\ are important outside
the galactic plane, where they control the increase in fluctuations with
$\left\vert Z\right\vert $, as shown in sectors E,F in Figure~\ref{fg9} and the
slope of curves in Figure~\ref{fg8NS}. These figures suggest that the collective velocity decreases with increasing $\left\vert Z\right\vert $\ and the
directions of the trajectories are becoming more random and probably less
circular. The further effect of $\sigma_{\Phi1}%
^{-}$ is due to the asymmetry of the distribution $V_{\Phi}$, which generates
the deceleration of collective rotation with increasing $\left\vert Z\right\vert $
according to Equation~(\ref{r13b}). This asymmetry is also manifested very clearly in
Figures~\ref{fg11} and~\ref{fg12} (see the comments in the last two paragraphs of
Section~\ref{cosida}).

(c) Our assumption that $\sigma_{\Phi}^{+}$ does not depend on $\left\vert
Z\right\vert $ can be verified by comparing $v_{l}$ distributions in the sectors
Q$_{1N}$,Q$_{3N}$,Q$_{1S}$, and Q$_{3S}$ in Figure~\ref{fg12} with the corresponding
distributions in sectors A and C in Figure~\ref{fg10}. This independence means
that in the analysed region,
\begin{equation}
\left\vert \mathbf{V}\left(  R,Z\right)  \right\vert \lesssim V_{0}%
+\sigma_{\Phi0}\approx250\,\mathrm{km/s};\qquad5\lesssim R\lesssim
13\mathrm{kpc},\qquad\left\vert Z\right\vert \lesssim3\mathrm{kpc.}\label{r20}%
\end{equation}
We showed that the 3D Monte Carlo model fits all studied sectors of the averaged kinematic data very well. Of course, its parameters may require further optimisation in more distant regions.

To conclude, the proposed statistical methods for calculating the local
velocity of the Sun, the average rotation velocity $V_{0}$, and generally the velocity $V_{G}(R,Z)$
at different positions in the MW can be useful for the analysis of the current and
future Gaia data releases. It is always important to be able to compare these
parameters obtained by different methods and input data samples.
Averaged, axisymmetric approximations of the MW
kinematics represented by the Monte Carlo simulation code can be useful in
validating axisymmetric dynamic models or determining the scale of local
kinematical substructures out of axial symmetry.

%\vspace{6pt}
%\authorcontributions{Methodology, P.Z.; Software, P.Z. and K.P.;  Formal Analysis, P.Z. and K.P.; Investigation, P.Z. and K.P.; Data Curation, K.P.; Writing---Original Draft Preparation, P.Z. All authors have read and agreed to the published version of the~manuscript.}%For research articles with several authors, a short paragraph specifying their individual contributions must be provided. The following statements should be used ``Conceptualization, X.X. and Y.Y.; methodology, X.X.; software, X.X.; validation, X.X., Y.Y. and Z.Z.; formal analysis, X.X.; investigation, X.X.; resources, X.X.; data curation, X.X.; writing---original draft preparation, X.X.; writing---review and editing, X.X.; visualization, X.X.; supervision, X.X.; project administration, X.X.; funding acquisition, Y.Y. All authors have read and agreed to the published version of the manuscript.'', please turn to the  \href{http://img.mdpi.org/data/contributor-role-instruction.pdf}{CRediT taxonomy} for the term explanation. Authorship must be limited to those who have contributed substantially to the work~reported.

%\funding{{This research received no external funding.}}%Please add: ``This research received no external funding'' or ``This research was funded by NAME OF FUNDER grant number XXX.'' and  and ``The APC was funded by XXX''. Check carefully that the details given are accurate and use the standard spelling of funding agency names at \url{https://search.crossref.org/funding}, any errors may affect your future funding.
%\section*{Acknowledgements}

%\dataavailability{
The data underlying this article were accessed from the European Space Agency (ESA) mission Gaia  (\url{https://www.cosmos.esa.int/gaia} ({accessed} on 20 January 2024)) and from publicly available cited references. The derived data generated in this research will be shared on reasonable request to the corresponding author.
%}%{Please refer to suggested Data Availability Statements in section “MDPI Research Data Policies” at \href{https://www.mdpi.com/ethics}{https://www.mdpi.com/ethics}}.  In this section, please provide details regarding where data supporting reported results can be found, including links to publicly archived datasets analyzed or generated during the study. Please refer to suggested Data Availability Statements in section “MDPI Research Data Policies” at https://www.mdpi.com/ethics. You might choose to exclude this statement if the study did not report any data.
%MDPI: If there are new data generated in this work, please consider to revise the "Data Availability Statements".

\vspace{-3mm}
\acknowledgments{%
This %MDPI: To AE: please confirm if the funding information in the Acknowledgments Section should be moved to the Funding Section. PZ: Done
work made use of data from the European Space Agency (ESA) mission
\textit{Gaia} (\url{https://www.cosmos.esa.int/gaia} ({accessed on 20 January 2024} %MDPI: Please add the access date (format: Date Month Year), e.g., accessed on 1 January 2020. PZ: Done
)), processed by the
\textit{Gaia} Data Processing and Analysis Consortium (DPAC,
\url{https://www.cosmos.esa.int/web/gaia/dpac/consortium} ({accessed} on 20 January 2024)). Funding for the
DPAC was provided by national institutions, in particular the
institutions participating in the \textit{Gaia} Multilateral Agreement.  We
are grateful to A.Kup\v{c}o for the critical reading of the manuscript and
valuable comments. We are also grateful to J. Grygar for his deep interest and
qualified comments and to O. Teryaev for helpful discussions in the early
stages of this work. Last but not least, we thank the academic editor and anonymous reviewers for their thorough reading of the manuscript and critical recommendations for improvement.}

%\conflictsofinterest{{The authors declare no conflicts of interest.}}%Declare conflicts of interest or state ``The authors declare no conflict of interest.'' Authors must identify and declare any personal circumstances or interest that may be perceived as inappropriately influencing the representation or interpretation of reported research results. Any role of the funders in the design of the study; in the collection, analyses or interpretation of data; in the writing of the manuscript, or in the decision to publish the results must be declared in this section. If there is no role, please state ``The funders had no role in the design of the study; in the collection, analyses, or interpretation of data; in the writing of the manuscript, or in the decision to publish the~results''.
%%%%%%%%%%%%%%%%%%%%%%%%%%%%%%%%%%%%%%%%%%
%\begin{adjustwidth}{-\extralength}{0cm}
%\printendnotes[custom] % Un-comment to print a list of endnotes

%\reftitle{References}

%=====================================

%\PublishersNote{}
%\end{adjustwidth}

\end{document}